\documentclass[structabstract]{aa}

\usepackage{url}
\usepackage[breaklinks=true]{hyperref}
\usepackage{twoopt}
\usepackage[english]{babel}          

\usepackage{natbib}
\bibpunct{(}{)}{;}{a}{}{,} 

\usepackage[utf8]{inputenc}
\usepackage[normalem]{ulem}
\usepackage{siunitx}
\usepackage{amsmath, amssymb}
\usepackage[farskip=0pt]{subfig}
\usepackage{color}

\usepackage{multirow,bigstrut,ctable}
\usepackage[normalem]{ulem}
\usepackage{rotating}
\usepackage[varg]{txfonts} 
\usepackage{threeparttable}

\def \kms         {km$\,$s$^{-1}$}

\def \deg         {\text{$^{\circ}$}}
\def \arcmin      {\text{$^\prime$}}
\def \arcsec      {\text{$^{\prime\prime}$}}

\def \mjybeam     {mJy\,beam$^{-1}$}

\def \mach        {\mathcal{M}}

\newcommand{\beam}[2]{{#1}\arcsec$\times${#2}\arcsec}

\begin{document}

\title{Reaching thermal noise at ultra-low radio frequencies}
\subtitle{ Toothbrush radio relic downstream of the shock front}

\titlerunning{The Toothbrush cluster at 58 MHz}

\author{F.~de~Gasperin\inst{1}
\and G. Brunetti\inst{5}
\and M. Br\"uggen\inst{1}
\and R. van Weeren\inst{2}
\and W. L. Williams\inst{2}
\and A. Botteon\inst{2}
\and V. Cuciti\inst{1}
\and T. J. Dijkema\inst{3}
\and H. Edler\inst{1}
\and M. Iacobelli\inst{3}
\and H. Kang\inst{4}
\and A. Offringa\inst{3}
\and E. Orr\'u\inst{3}
\and R. Pizzo\inst{3}
\and D. Rafferty\inst{1}
\and H. R\"ottgering\inst{2}
\and T. Shimwell\inst{2,3}}
\authorrunning{F.~de~Gasperin et al.}

\institute{ Hamburger Sternwarte, Universit\"at Hamburg, Gojenbergsweg 112, 21029, Hamburg, Germany, \email{fdg@hs.uni-hamburg.de}
\and Leiden Observatory, Leiden University, P.O.Box 9513, NL-2300 RA, Leiden, The Netherlands
\and ASTRON, Netherlands Institute for Radio Astronomy, PO Box 2, NL-7990 AA Dwingeloo, the Netherlands
\and Department of Earth Sciences, Pusan National University, Busan 46241, Korea
\and INAF - Istituto di Radioastronomia, via P. Gobetti 101, Bologna, Italy }

\date{Received ... / Accepted ...}

\abstract
{Ultra-low frequency observations ($<100$ MHz) are particularly challenging because they are usually performed in a low signal-to-noise ratio regime due to the high sky temperature and because of ionospheric disturbances whose effects are inversely proportional to the observing frequency. Nonetheless, these observations are crucial for studying the emission from low-energy populations of cosmic rays.}
{We aim to obtain the first thermal-noise limited ($\sim 1.5$~\mjybeam{}) deep continuum radio map using the  Low Frequency Array's Low Band Antenna (LOFAR LBA) system. Our demonstration observation targeted the galaxy cluster RX J0603.3+4214 (known as the Toothbrush cluster). We used the resulting ultra-low frequency (39 -- 78 MHz) image to study cosmic-ray acceleration and evolution in the post shock region considering the presence of a radio halo.}
{We describe the data reduction we used to calibrate LOFAR LBA observations. The resulting image was combined with observations at higher frequencies (LOFAR 150 MHz and VLA 1500 MHz) to extract spectral information.}
{We obtained the first thermal-noise limited image from an observation carried out with the LOFAR LBA system using all Dutch stations at a central frequency of 58 MHz. With eight hours of data, we reached an rms noise of 1.3~\mjybeam{} at a resolution of \beam{18}{11}.}
{The procedure we developed is an important step towards routine high-fidelity imaging with the LOFAR LBA. The analysis of the radio spectra shows that the radio relic extends to distances of 800 kpc downstream from the shock front, larger than what is allowed by electron cooling time. Furthermore, the shock wave started accelerating electrons already at a projected distance of $<300$ kpc from the crossing point of the two clusters. These results may be explained by electrons being re-accelerated downstream by background turbulence, possibly combined with projection effects with respect to the radio halo.}


\maketitle

\section{Introduction}
\label{sec:introduction}

The Low Frequency Array \citep[LOFAR;][]{VanHaarlem2013} is currently the only instrument capable of conducting sensitive high-resolution observations at ultra-low frequencies ($<100$~MHz). Such observations make use of the Low Band Antenna (LBA) system, which covers the frequency range of 10 -- 90 MHz. The imaging capabilities of the LBA will remain unique even into the era of the Square Kilometer Array \citep[SKA;]{Dewdney2013}, which is currently planned to observe above 50~MHz and at lower resolution.

Despite LBA operating routinely since 2012, very few continuum images taken with the instrument have been published \citep[e.g.][]{vanWeeren2012f} because sufficiently high-fidelity images have largely remained elusive. This is primarily because of the effect caused by the varying ionosphere over the array \citep{Mevius2016, deGasperin2018a}. The ionosphere is a magnetised plasma that surrounds our planet and, at first order, it imprints a variable delay in the incoming radio waves that is proportional to the number of free electrons (known as the Total Electron Content, TEC) along the observing direction. The delay is highly direction-dependent and it becomes stronger at a lower frequency, following a $\propto 1/\nu$ proportionality. Higher order effects such as Faraday rotation ($\propto 1/\nu^2$) are also non-negligible when observing at these frequencies.

Radio sources tend to have increasing flux density with decreasing low frequency. The slope of the radio spectra is on average $-0.7-0.8$ for bright sources \citep[e.g.][]{Vollmer2010, ki08, deGasperin2018}. However, with decreasing frequency, the sky temperature also increases dramatically ($T_B \propto \nu^{-\beta}$ with $\beta=2-3$). Consequently, ultra-low frequency observations are typically in a low signal-to-noise ratio (S/N) regime.

\begin{figure*}[htb!]
\centering
 \includegraphics[width=.49\textwidth]{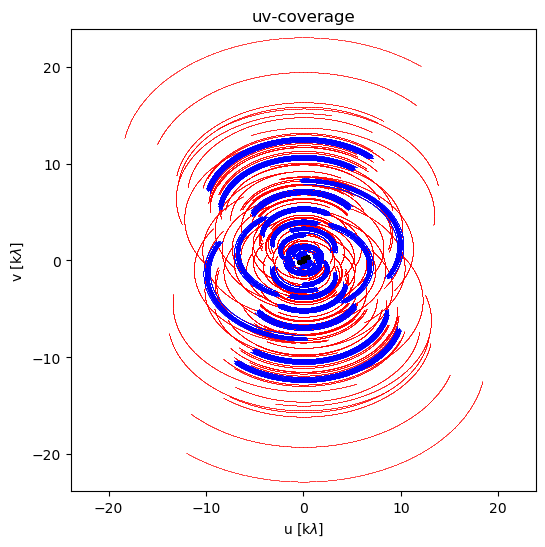}
 \includegraphics[width=.49\textwidth]{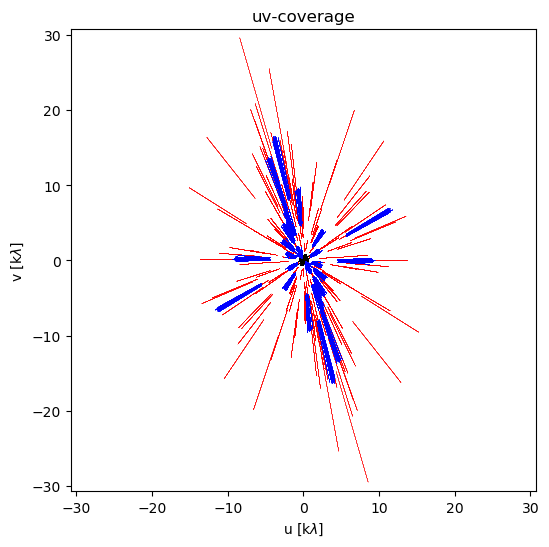}
 \caption{$uv$-coverage for target field of the observation described in the text. Left panel is the coverage for a single channel at the central frequency of 58 MHz. Right panel is the instantaneous coverage across the entire band. Baseline combinations with different stations are colour-coded: black for core-core baselines, blue for core-remote, and red for remote-remote baselines.}
 \label{fig:uv}
\end{figure*}

An eight-hour observation using the LOFAR LBA system covering 30 -- 74 MHz, without international stations, should result in an rms noise level of $\sim 1 - 1.5$~\mjybeam{} with a resolution of 15\arcsec{} \citep{VanHaarlem2013}. The precise value in the range is set by the projected size of the LOFAR stations, which is dependent upon the target elevation throughout the observation. Simulations that take into account realistic noise, the effect of the bandpass and of the dipole beam for an observation like the one discussed in this paper provided a thermal noise of 1.1~\mjybeam (Edler et al. in prep.). 

Thermal noise calibration and imaging procedures for LOFAR do exist \citep{vanWeeren2016b, Shimwell2019}, however these strategies are tailored to the LOFAR High Band Antenna (HBA), which operates in the frequency range of $120-240$ MHz and in a higher S/N regime. In this paper, we outline a calibration and imaging strategy that is tailored to optimise the image fidelity of the LBA.

\subsection{The Toothbrush cluster}

RX J0603.3+4214, the Toothbrush, is a merging galaxy cluster at $z=0.225$ \citep{Dawson2015}. This system hosts three radio relics: two smaller ones on the southern side and a larger one ($\sim2$ Mpc) in the north. The latter is one of the largest and brightest radio relics known \citep{vanWeeren2012e, vanWeeren2016, Rajpurohit2018, Rajpurohit2020}. Between these relics, and connected to them, there is a radio halo. The spectral shape of the northern relic goes from flat to steep moving inwards towards the cluster centre, while the integrated spectral shape is a straight power-law from 74 MHz up to 8 GHz with slope $\alpha = -1.10\pm0.02$. Downstream of the shock front, the plasma that was energised by the shocks responsible for the radio relics merges with emission from the radio halo region, possibly contributing mildly energetic cosmic rays to that region. The open question that we try to tackle in
this paper considers the processes that are at work far downstream of the shock front.

From the radio spectra and assuming standard diffuse shock acceleration (DSA), the northern relic should trace a moderately strong shock wave propagating northwards with a Mach number of $\mach = 3.3 - 4.6$. Using numerical simulations, \cite{Bruggen2012} suggested a triple-merger scenario, with a main merger oriented north-south and a second merger of a smaller structure occurring on the south-west axis. With XMM-Newton observations, \cite{Ogrean2013} showed that the main merger axis is in fact oriented north-south, but they were unable to find evidence of smaller substructures. Chandra observations were analysed in \citep{vanWeeren2016}, with the reported detection of surface brightness discontinuities compatible with the presence of shock fronts both at the northern and southern edges of the cluster. Both XMM-Newton and Chandra observations find only marginal evidence of a weak ($\mathcal M \approx 1.2$) shock at the external edge of the main radio relic. \cite{Dawson2015} identified a number of substructures in the cluster picturing a rather complex merger scenario.

The paper is organised as follows. We describe the LBA observations in Sec.~\ref{sec:observations},
and the data reduction pipeline we implemented in Sec.~\ref{sec:calibration}  In Sec.~\ref{sec:results}, we present our results for the Toothbrush cluster. The discussion and conclusions are given in Sec.~\ref{sec:discussion} and \ref{sec:conclusions}. Throughout this paper, we adopt a fiducial $\Lambda$CDM cosmology with $H_0 = 70\rm\ km\ s^{-1}\ Mpc^{-1}$, $\Omega_m = 0.3,$ and $\Omega_\Lambda = 0.7$. At the redshift of the Toothbrush cluster ($z\approx0.225$), we have 1\arcsec = 3.614 kpc. Unless otherwise specified errors are at $1\sigma$. The spectral index is defined as: $F_{\nu} \propto \nu^\alpha$, where $F_\nu$ is the flux density.

\section{Observations}
\label{sec:observations}

Our eight-hour observation was performed using the LOFAR LBA system in the frequency range of 39 -- 78 MHz, where the antennas are most sensitive. We used 24 Core Stations and 13 Remote Stations. The correlated data had an integration time of 1~s and a frequency resolution of 64 channels per 0.192~MHz SubBand (SB). The $uv$-coverage for the target field is in Fig.~\ref{fig:uv}.

The observation was carried out in multi-beam mode, with one beam continuously pointing at the calibrator (3C147) and one beam continuously pointing at the target (RX J0603.3+4214). This observing scheme can be adopted when observing with the LBA system because the beams are formed digitally and they can be arbitrarily placed on the sky. Using one beam to monitor a calibrator allows us to track the instrumental systematic effects along the entire observations using the high S/N given by the calibrator.

We used the LBA\_OUTER antenna setup, where only the LBA dipoles in the outer half of each LOFAR station record data and is generally used to avoid cross-talk that effects the inner most dipoles more severely and to contain the size of the primary beam. At the declination of the Toothbrush cluster, the primary beam FWHM is $\sim 4\deg$ with the maximum side lobes occurring at $\sim 7\deg$ from the pointing centre. We note that as the fractional bandwidth of the LBA system is large (70\% in this case), the primary beam FWHM increases approximately from 3\deg{} to 5\deg across the band.

\begin{table}
\centering
\begin{threeparttable}
\begin{tabular}{ll}
Calibrator & 3C147\bigstrut[t]\\
Target & RX J0603.3+4214 (Toothbrush)\\
Observing time & 2019-04-17, 13:00$\ -\ $21:00\\
Time resolution\tnote{a} & 1 s $\rightarrow$ 4 s\\
Frequency range & 39 -- 78 MHz\\
Frequency resolution\tnote{a,b} & 64 ch/SB $\rightarrow$ 4 ch/SB \\
System & LOFAR LBA\_OUTER \\
Correlations & Full Stokes \\
\end{tabular}
\begin{tablenotes}
    \item[a] Before and after initial averaging.
    \item[b] SubBand (SB) bandwidth: 0.192 MHz.
\end{tablenotes}
\end{threeparttable}
\caption{Observation logs}\label{tab:1}
\end{table}

\begin{figure}
\centering
\includegraphics[width=.4\textwidth]{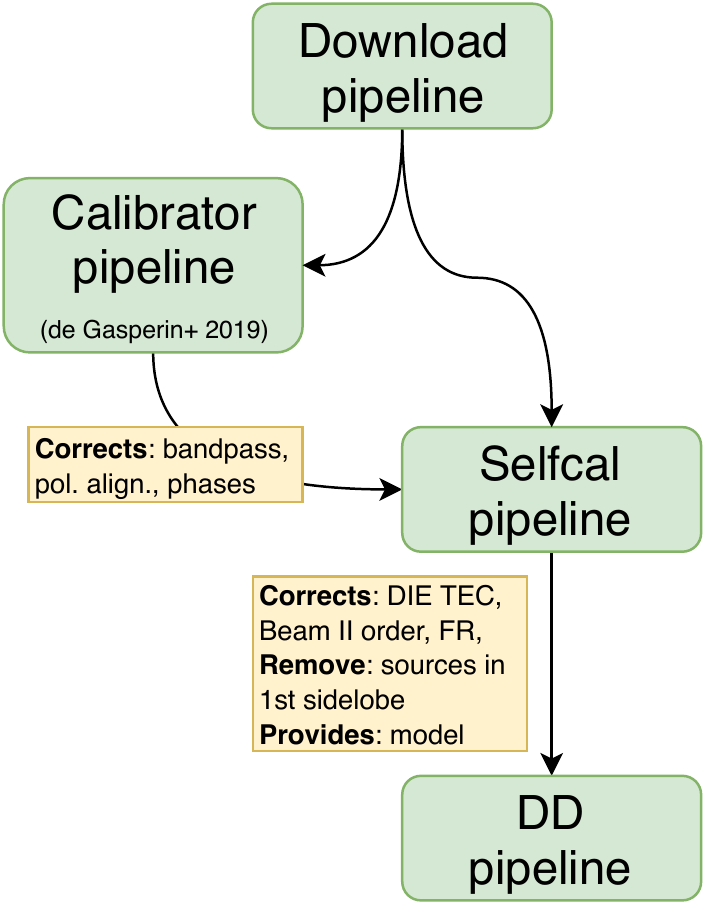}
\caption{Scheme that summarises the steps of our calibration pipeline that starts after the pre-processing pipeline run by the observatory. In green, we show the four macro-steps described in detail in the text or in \citet{deGasperin2019}. In yellow, we show the effect that one step has on another.}\label{fig:pipeline}
\end{figure}

\section{Calibration}
\label{sec:calibration}

Figure~\ref{fig:pipeline} outlines the four major steps of our data reduction pipeline. In the following sections, we describe each part of the pipeline in detail. When needed, we use the radio interferometer measurement equation (RIME) formalism to describe LOFAR systematic effects. The formalism is described in detail in the first two papers of `Revisiting the radio interferometer measurement equations' \citep{Smirnov2011a, Smirnov2011b}.

\subsection{Observatory pre-processing and download}
\label{sec:download}

The correlated data (1 s and 64 ch/SB resolution) for an eight-hour LBA observation with 37 stations has a data volume of 10 TB. This high resolution is required for Radio Frequency Interference (RFI) identification and for the subtraction of the bright sources seen through the far side-lobes \citep{deGasperin2020}. Both steps are offered as part of the ASTRON radio observatory Pre-Processing pipeline. The first makes use of AOflagger \citep{Offringa2010, Offringa2012} and the second removes contaminating sources via the Demix procedure \citep{VanderTol2009}. In our case we demixed both Cygnus A and Cassiopeia A. After these procedures, the observatory averaged the data to 4 s and 4 ch/SB and further reduced by compression \citep{Offringa2016}, reaching a size of 260 GB, half for the calibrator and half for the target. These data are then stored on the Long Term Archive \citep[LTA;][]{Belikov2011} and the larger, rawer data are discarded. The LTA data were then downloaded to local compute facilities for further processing.

\subsection{Calibrator}
\label{sec:calibrator}

The calibrator beam is pointed towards 3C147. Due to the turnover of its spectrum at $\sim 150$ MHz, this source is sub-optimal compared to other calibrators, especially at ultra-low frequencies \citep{Scaife2012, deGasperin2019}. Nonetheless, we managed to obtain good calibration solutions down to $\sim 45$ MHz. Below that frequency, the flux density of the calibrator source becomes critically low yielding data loss because of the poor S/N. We therefore restricted our analysis from both calibrator and target to data above 45 MHz.

\begin{figure}
\centering
 \includegraphics[width=.49\textwidth]{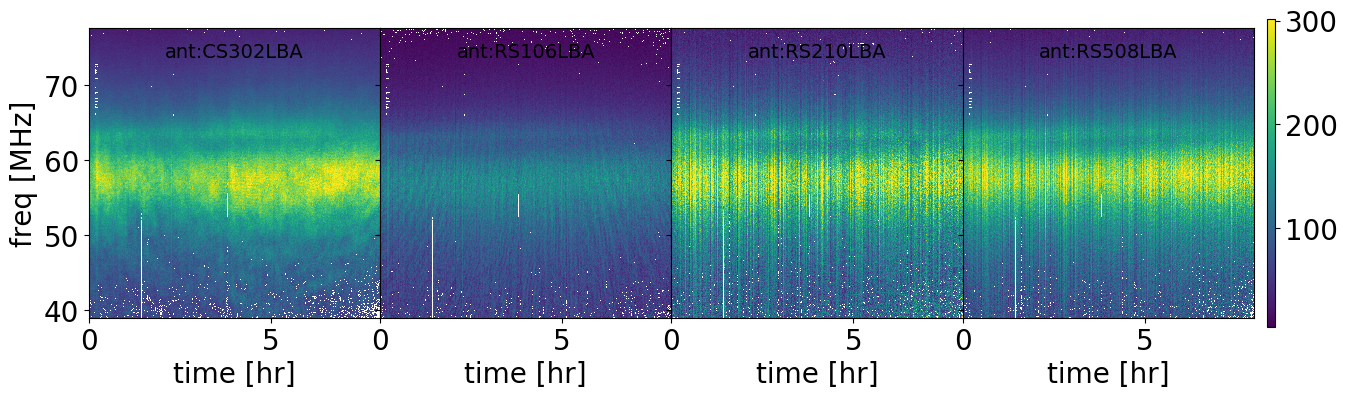}\\
 \includegraphics[width=.49\textwidth]{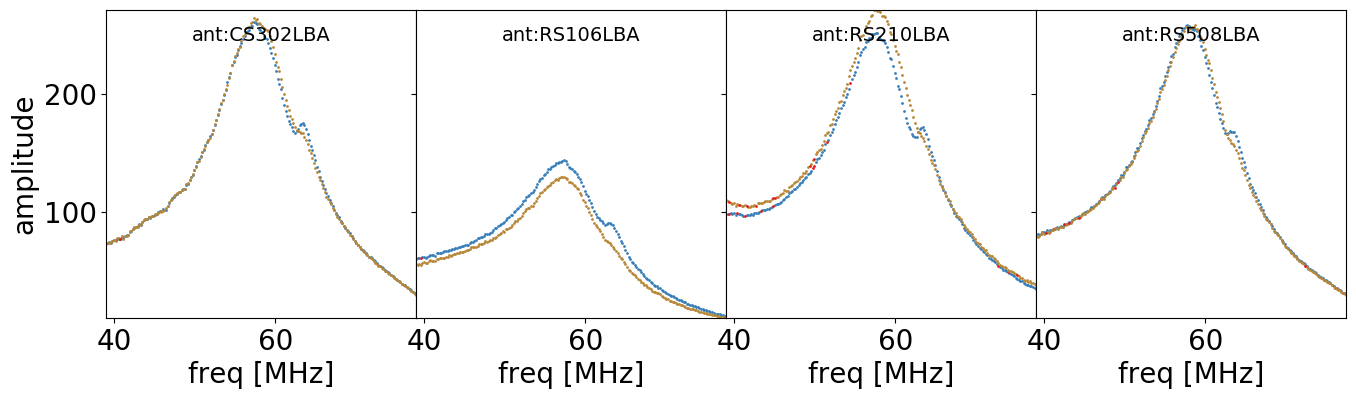}
 \caption{Top panel: Amplitude solutions for four different stations (CS302, RS106, RS210, and RS508) plotted as a function of observing time (x-axis) and frequency (y-axis). The bottom panel shows the time-averaged amplitude solutions that encode the instrument frequency response and that are transferred to the target field.}
 \label{fig:sol_cal1}
\end{figure}

\begin{figure}
\centering
 \includegraphics[width=.49\textwidth]{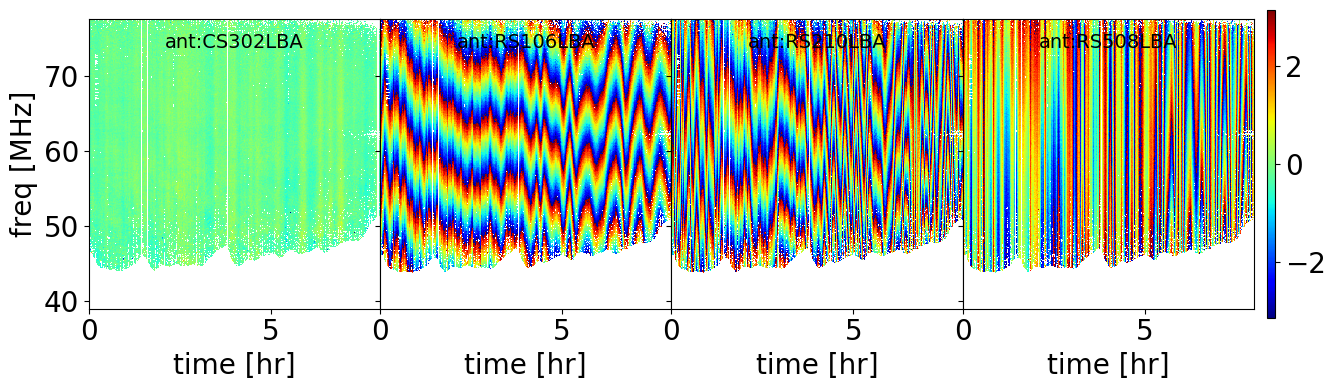}
 \includegraphics[width=.49\textwidth]{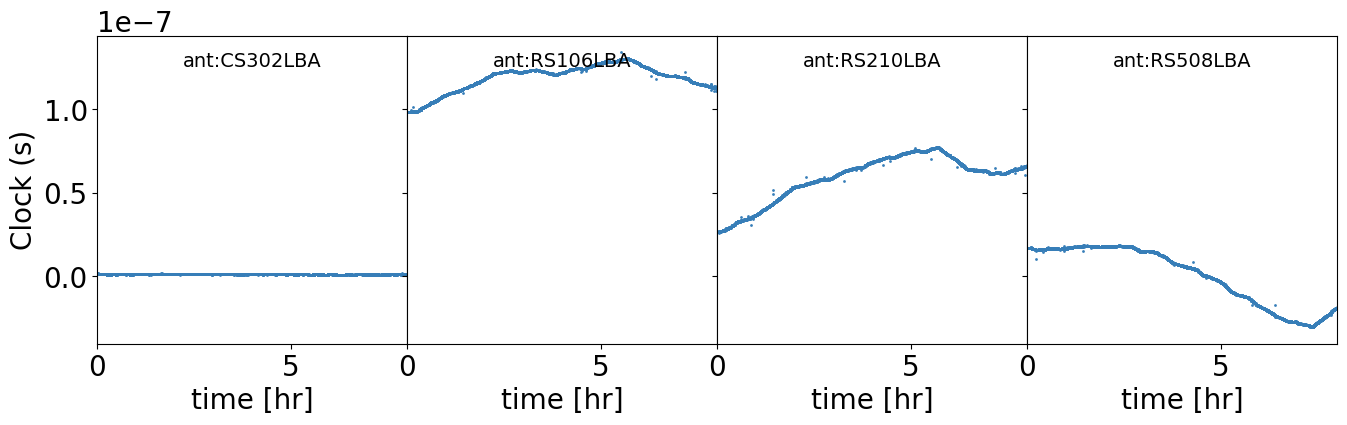}
 \includegraphics[width=.49\textwidth]{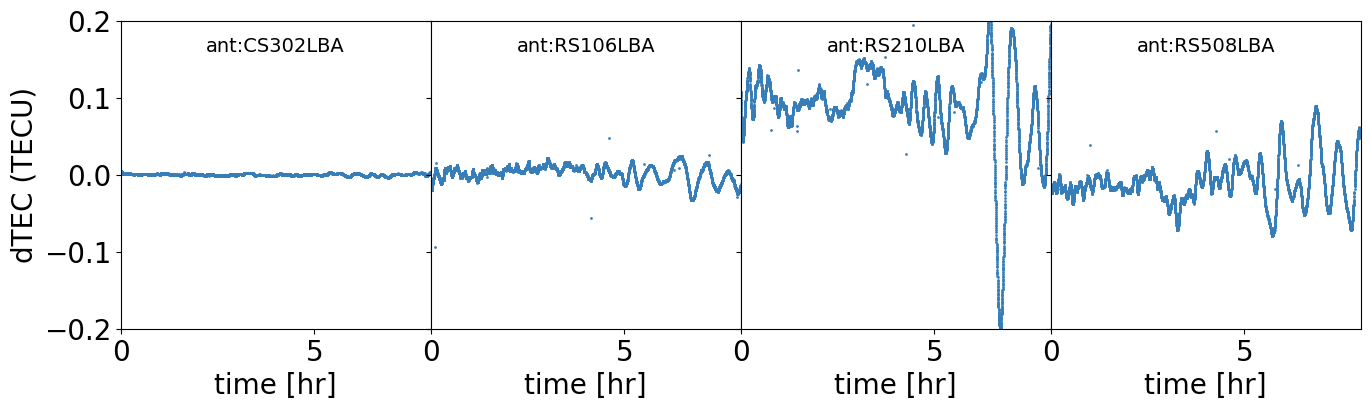}
 \includegraphics[width=.49\textwidth]{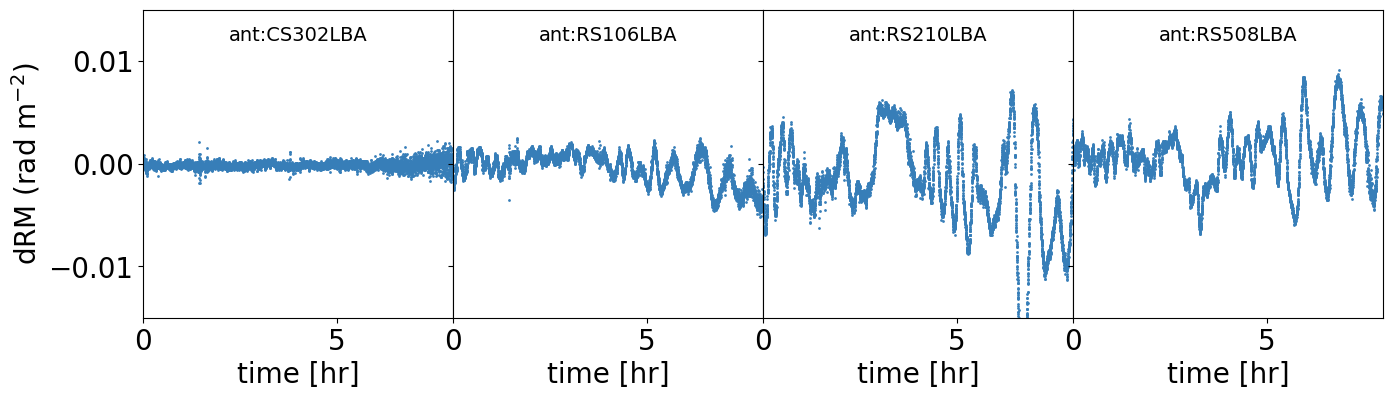}
 \caption{Top panel: Phase solutions in radians for four different stations (CS302, RS106, RS210, and RS508) plotted as a function of observing time (x-axis) and frequency (y-axis), referenced to station CS002 at the array centre. Second to fourth panels: clock, differential TEC, and differential Faraday rotation solutions extracted during the calibration process.}
 \label{fig:sol_cal2}
\end{figure}

The data reduction of 3C147 followed closely that described by \cite{deGasperin2019}. The procedure is pipelined in the so-called "Prefactor 3" and freely available to download\footnote{\url{https://github.com/lofar-astron/prefactor}}. Here we report some inspection plots that summarise the results and give an impression of the data quality. In Fig.~\ref{fig:sol_cal1}, we show the amplitude solutions used to extract the median bandpass over the duration of the observation. Some data in the lower frequency part of the bandpass are flagged due to the low S/N. The second half of the calibrator observation is generally noisier because the source is at lower elevation. In Fig.~\ref{fig:sol_cal2}, we show the phase solutions for the XX correlation. Several channels below 45 MHz were removed because the data were noisy. The bottom three diagnostic plots show the estimated clock drift, the differential TEC, and differential Faraday rotation (FR) with respect to the superterp stations\footnote{The six innermost LOFAR stations are packed within 1 km$^2$ and are collectively called the ``superterp''.}. The later effects strongly correlate because FR is generated by a combination of varying TEC and spatially varying Earth magnetic field \citep{deGasperin2018a}.

\begin{figure}
\centering
\includegraphics[width=.5\textwidth]{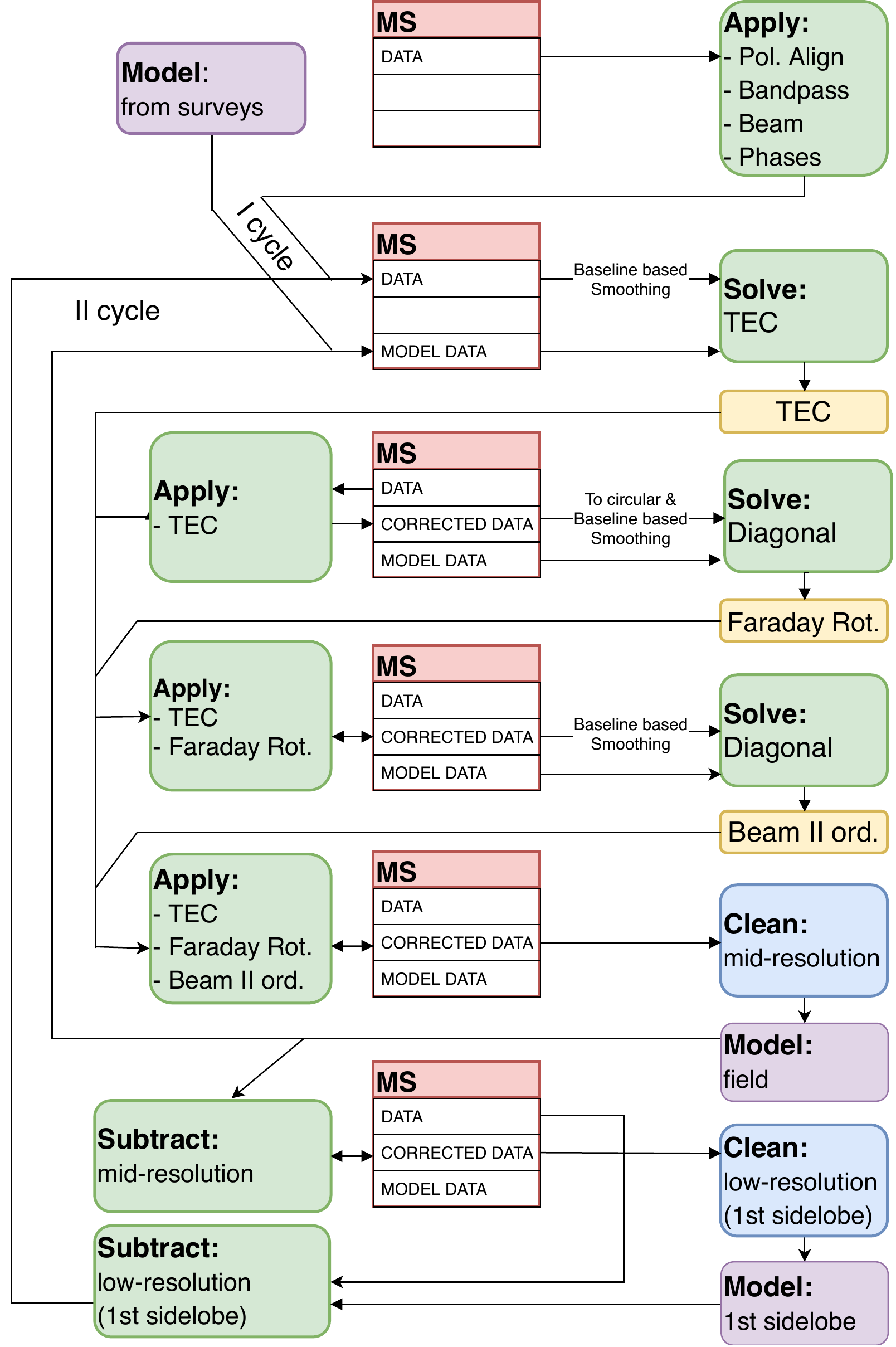}
\caption{Scheme showing the steps of the self-calibration pipeline. Steps indicated in green are solve, apply, and subtraction steps and are carried out with DPPP \citep{vanDiepen2018}. Steps shown in yellow consist of solutions manipulations and are carried out by LoSoTo \citep{deGasperin2019}. Blue steps represent imaging made with WSClean \citep{Offringa2014} and purple boxes are source models. Each solve step has an input data column and also uses data from the model. Each apply and subtract step has an input data column and an output data column.}\label{fig:pipeline_self}
\end{figure}

A total of three solution sets were collected from the calibrator run and used to correct the target data. Firstly, the polarisation alignment phase solutions (a diagonal delay Jones matrix) are transferred and applied. These differential delays are introduced by approximations or errors in the LOFAR station calibration table \citep{deGasperin2019} and need to be applied before the beam correction, which is a full-Jones matrix. Secondly, the bandpass and the phases (including ionospheric delay and clock corrections) were transferred. The bandpass and the clock are instrumental (direction-independent) errors, therefore their effect is the same for the calibrator and the target direction. After this correction, these effects are removed from the target data. As the ionosphere is direction-dependent, applying phase solutions extracted from the calibrator data to the target field will corrupt the data with the ionospheric phase effect in the calibrator direction, that is, after applying these solutions to the target field for each antenna the ionospheric effects now in the data are differential with respect to those in the direction of the calibrator. Ideally, we would not apply these phase solutions and would instead separate the ionospheric systematic effects from the instrumental ones (clock delays) and only apply the direction independent instrumental effects. To separate these effects, we can exploit the different frequency dependency of the effects and disentangle their contribution to the phase solutions. This process is known as clock/TEC separation \citep{Mevius2016}. However, in a poor S/N regime, the effects are not robustly separated one from the other and hence, based on our experience, it is best to transfer the full phase solutions and deal with a differential ionosphere. As a last correction, the LOFAR theoretical element-beam was applied. These four correction steps are summarised in the top-right 'apply' block in Fig.~\ref{fig:pipeline_self}. Finally, all SBs are combined in one MeasurementSet and this is split up in time-chunks of length of one hour so that subsequent, more computationally expensive, processing is easily conducted in parallel.

\begin{figure}
\centering
 \includegraphics[width=.49\textwidth]{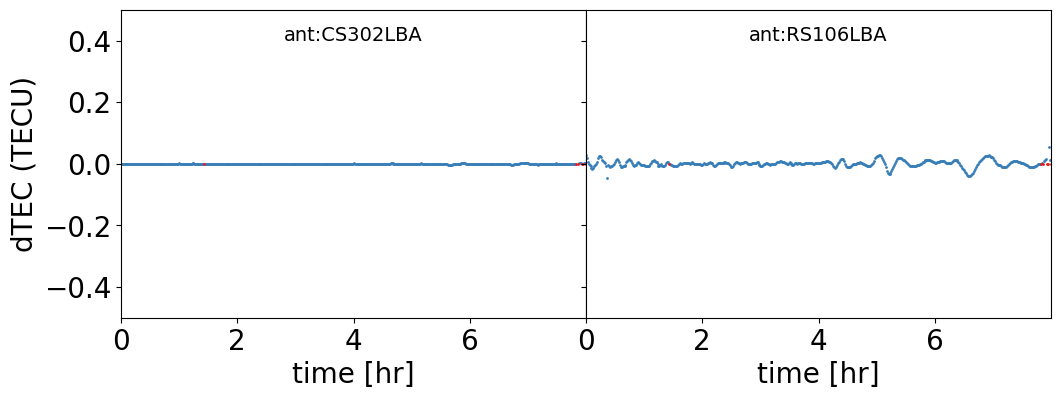}
  \includegraphics[width=.49\textwidth]{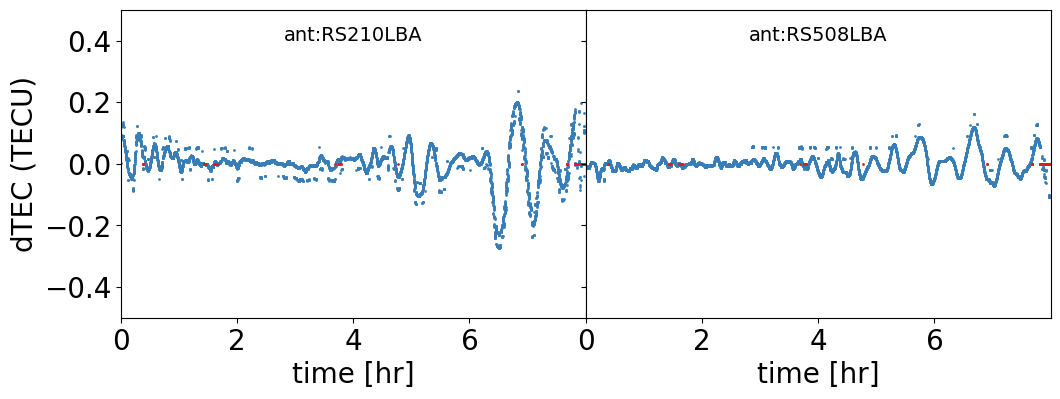}
 \caption{Differential TEC (dTEC) solutions (referenced to CS002) for four stations (CS302, RS106, RS210, and RS508) in the Toothbrush cluster direction. CS302 and RS106 are relatively close to the reference station, while the other two stations are farther away. Due to direction-dependent effects and the incompleteness of the skymodel, some errors in the estimation of the dTEC are present. Because of phase periodicity, the chi-square in TEC space has multiple local minima that results in discrete jumps of the dTEC values we recovered.}
 \label{fig:sol_self1}
\end{figure}

\begin{figure}
\centering
 \includegraphics[width=.49\textwidth]{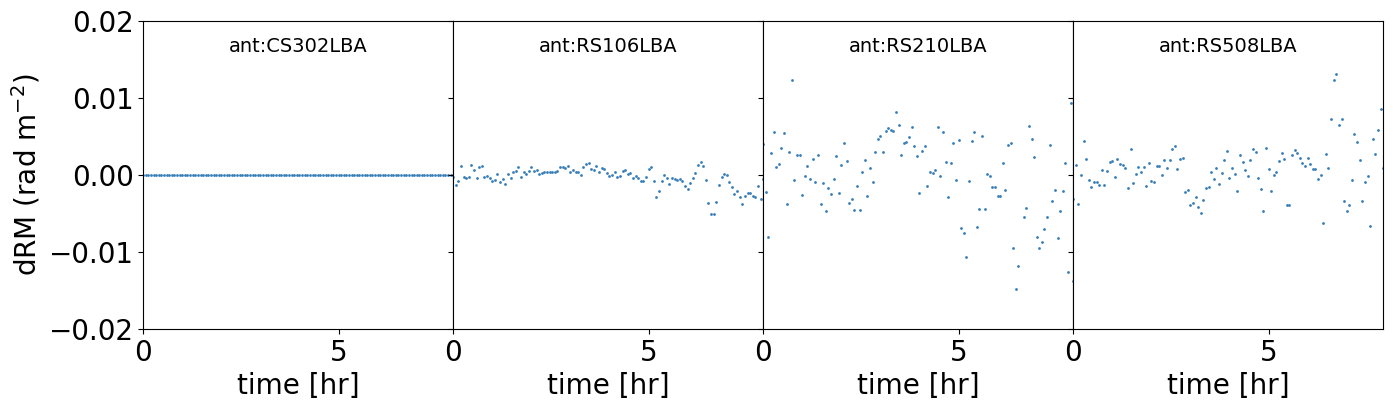}\\
 \includegraphics[width=.49\textwidth]{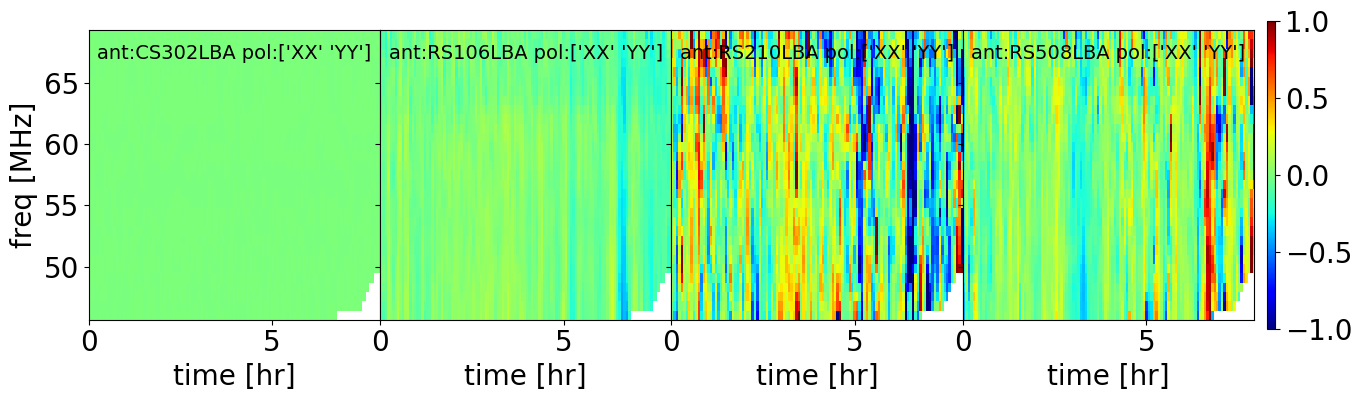}
 \includegraphics[width=.49\textwidth]{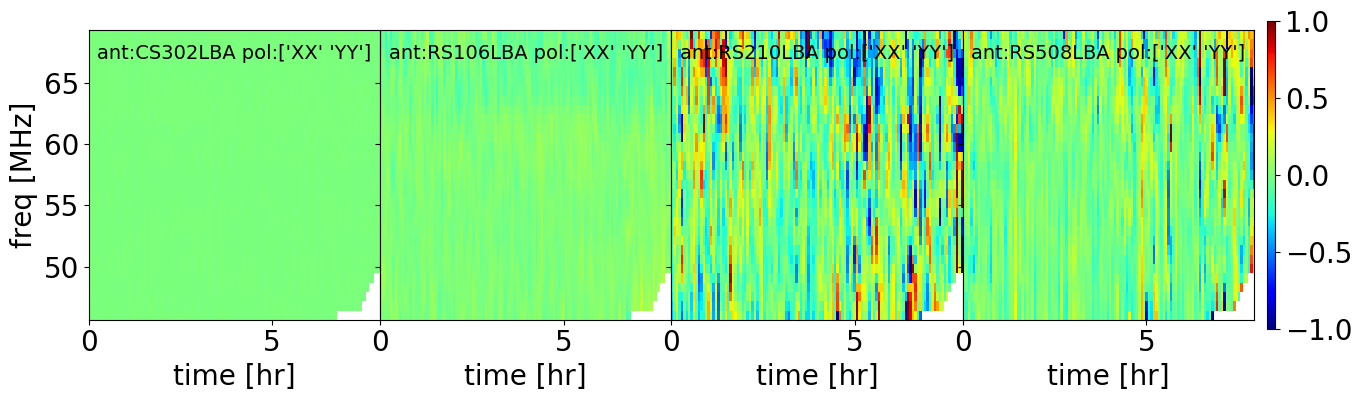}\\
 \caption{Top panel: Differential RM solutions (referenced to CS002) for the Toothbrush field for CS302, RS106, RS210, and RS508. Middle panel: Differential (RR - LL) phase solutions in circular basis expressed in radians plotted as a function of observing time (x-axis) and frequency (y-axis) and also referenced to CS002. Bottom panel: Same as middle panel but after the removal of the estimated RM effect.}
 \label{fig:sol_self2}
\end{figure}

\begin{figure}
\centering
 \includegraphics[width=.24\textwidth]{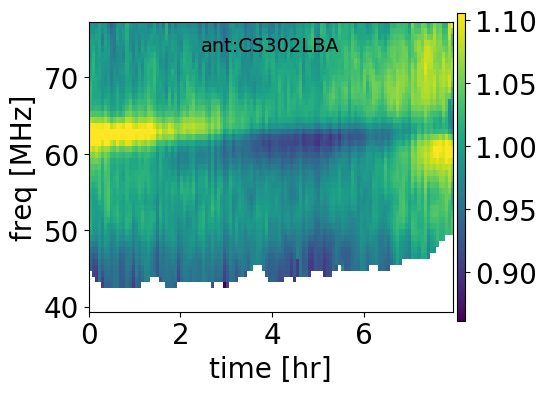}
 \includegraphics[width=.24\textwidth]{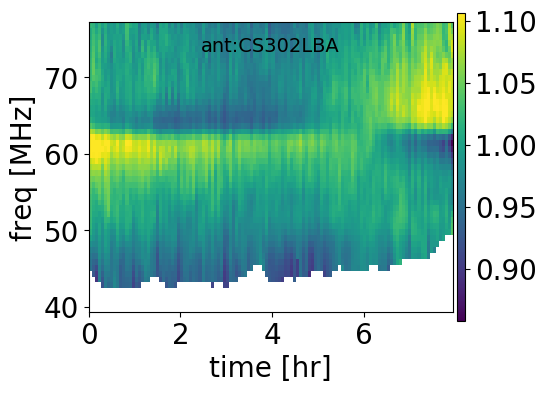} 
 \caption{Antenna-averaged normalised amplitude solutions in the direction of Toothbrush cluster plotted as a function of observing time (x-axis) and frequency (y-axis) for the X (left), and Y (right) polarisation. The bandpass has been already removed and here the dominant effects are unmodelled structures in the theoretical beam.}
 \label{fig:sol_self3}
\end{figure}

\subsection{Self-calibration}
\label{sec:selfcal}

Even after the solutions from the calibrator field have been applied to the target field, the target data still require extensive calibration to correct for the differential ionospheric effects. We performed an initial phase-only calibration step against a model of the field within the full width at half maximum (FWHM) constructed from a from the available surveys. Currently we use a combination of TGSS \citep{Intema2017}, NVSS \citep{Condon1998}, WENSS \citep{Rengelink1997}, and VLSS \citep{Lane2014}; so that we estimated the spectral index, when possible up to the second order, and extrapolated the flux density of each source to the LBA frequency range. Each source with an absolute flux density below 1 Jy at 60 MHz was discarded to reduce the model size. In this process, only a few per cent of the total flux density is lost. The model was then converted into visibilities accounting for the reduction in intensity due to the primary beam.

The process then proceeds as outlined in Fig.~\ref{fig:pipeline_self}. The data visibilities are Gaussian-smoothed in time and frequency with a kernel size that decreases with baseline length as described in \cite{deGasperin2019}. This increases the S/N in the visibilities by enforcing time and frequency coherency of the data on the short baselines. In practice, this is a workaround for the solver (DPPP) not being able to have an adaptive cell-size (in time and frequency) that is baseline length-dependent. Then, by calibrating our visibilities against the predicted model visibilities, we estimate a direction independent (field-averaged), TEC for all stations within 10 km from the Superterp by solving with time intervals of one minute and forcing the six stations in the Superterp to have a single TEC value. This assumption is made to increase the SNR. The solutions obtained were then pre-applied to the data before we estimated TEC solutions every four seconds for the other, more distant, stations. The TEC estimations are obtained with a phase solver where a TEC-delay $\propto\nu^{-1}$ is fit on scalar phase solutions each cycle until convergence. We plot the TEC solutions for some representative stations in Fig.~\ref{fig:sol_self1}.

After correcting the data for both sets of estimated TEC values, we converted the dataset to circular polarisation basis where Faraday rotation is easier to estimate and we solve for a diagonal Jones matrix. To increase the S/N during this step we: 1) find solutions on long intervals of four minutes; 2) force all stations within 10 km to have the same solutions; and 3) ensure that at each solver cycle \citep{Mitchell2008, Salvini2014a}, each solution set is smoothed in frequency with a Gaussian with $\sigma=3$~MHz (Offringa et al. in prep.). After these calibration solutions are derived, we use the differential values between the right- and the left-handed circular polarisation phase solutions to fit the rotation measure, assuming a frequency dependency of $\nu^{-2}$. We show the result of this procedure in Fig.~\ref{fig:sol_self2}.

At this point, the original (linear basis) dataset is corrected for Faraday Rotation by applying the solutions obtained from the circular polarisation basis dataset. Then we perform a second amplitude and phase diagonal solve, with the same strategy of the previous one. This time refining the amplitudes of the data. All core stations show the presence of amplitude errors of $\sim 10\%$ with very similar frequency-time structure. These are due to imperfect beam models, concentrated in the frequency range of $60-65$~MHz where the amplitude bandpass has a notch likely due to the edge of the dipole wire which forms a loop used to keep the wire in position. The size of the loop can vary from dipole to dipole and in certain cases can be wet, modifying the dipole theoretical response (M.J. Norden, priv. comm.). This second-order beam systematic effect is then averaged across all core stations to increase the S/N and after normalisation, it is finally applied to both core and remote stations (see Fig.~\ref{fig:sol_self3}). The correction can arrive to $\sim10\%$ in amplitude with the largest variations around the unmodelled wiggle that is also clearly visible in the calibrator bandpass around 62 MHz (see Fig.\ref{fig:sol_cal1}). Remote stations are excluded from the averaging as their amplitude solutions, at this stage, are affected by decorrelation caused by ionospheric disturbances.

\begin{figure}
\centering
 \includegraphics[width=.5\textwidth]{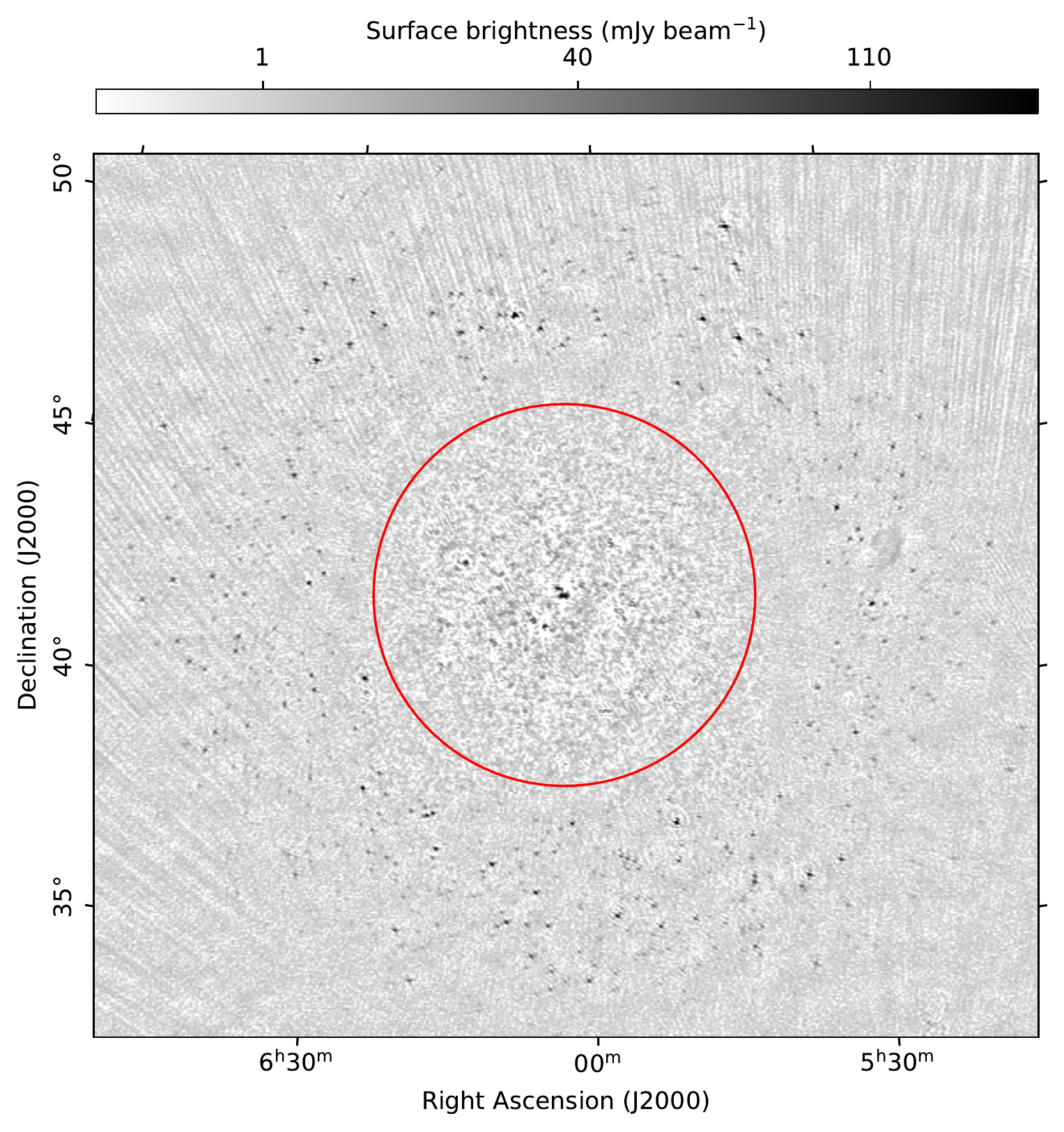}
 \caption{Low-resolution (beam: \beam{209}{144}) image of the Toothbrush cluster field after subtraction of mid-resolution sources in the main beam. The only visible sources are those present in the first side-lobe and some extended emission that was not fully subtracted from the high-resolution map. The red circle is an estimation of the first null at mid-frequency ($\approx 2 \cdot {\rm FWHM}$). Sources outside the circle are subtracted at the end of the first selfcal cycle. Stripes in the background are due to Taurus~A, which is at a distance of 22\deg{} and which was not demixed.}
 \label{fig:map_sidelobe}
\end{figure}

After the correction of the beam second-order effect, we produce a first wide-field image just using data in the $uv$-range 0 to 4500 lambda. This $uv$-range is used to remove stations that are the most affected by the, still uncorrected, direction dependent ionospheric corruptions. The resulting image covers the central $8\deg \times 8\deg$ (up to roughly the first null of the primary beam) and has a resolution of about 40\arcsec. We then subtract from the data all the sources detected and perform a second imaging step to capture sources in the first side-lobe, this time using only baselines shorter than 5 km to allow for a large pixel size and a large image size. Here, we image an area $25\deg \times 25\deg$ at the resolution of $\sim 3\arcmin$. For this field, the total apparent flux density of the sources detected in the first side lobe is $\sim 30$~Jy. By comparison, the combined apparent flux density of the sources in the main beam was 164 Jy. We isolate all sources outside the null at mid-frequency (see Fig.~\ref{fig:map_sidelobe}) and subtract them from the original visibilities after corrupting their signal with all systematic effects isolated so far.

\begin{figure}
\centering
 \includegraphics[width=.5\textwidth]{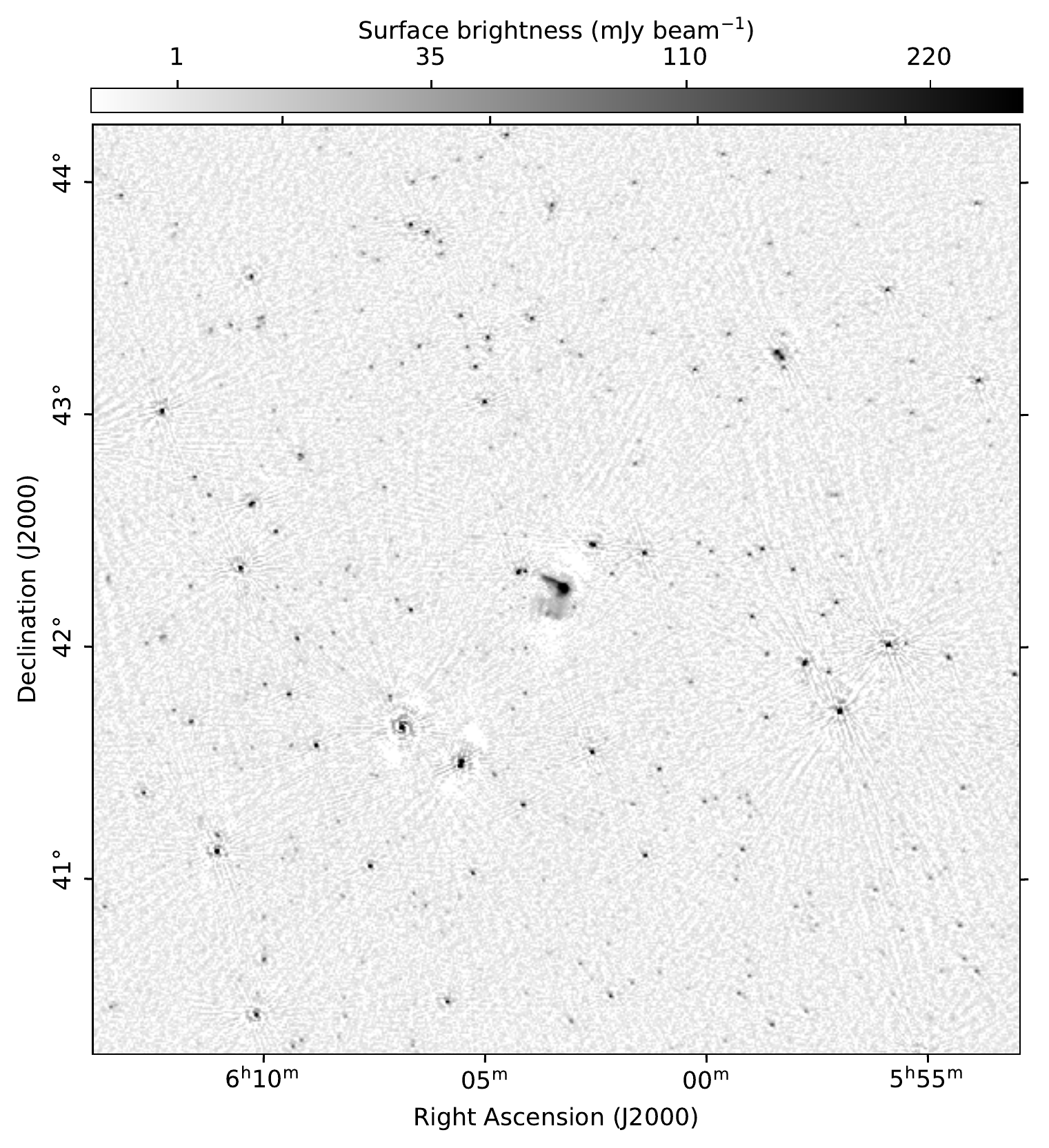}
 \caption{Radio map produced at the end of the self-calibration cycles. The rms noise is $\sim 2$~\mjybeam{} at a resolution of \beam{37}{29}. Residual errors around bright sources are due to the still-uncorrected direction dependent ionospheric corruptions.}
 \label{fig:map_die}
\end{figure}

We then start a second self-calibration cycle using an improved model obtained from the first mid-resolution imaging and a dataset with sources in the first side-lobe subtracted. We repeat the same calibration steps outlined above until we obtain a second image ($\sim 40\arcsec$ resolution), that is the final result of the self-calibration pipeline (Fig.~\ref{fig:map_die}).

\subsection{Direction-dependent calibration}
\label{sec:ddecal}

\begin{figure*}[!ht]
\centering
 \includegraphics[width=\textwidth]{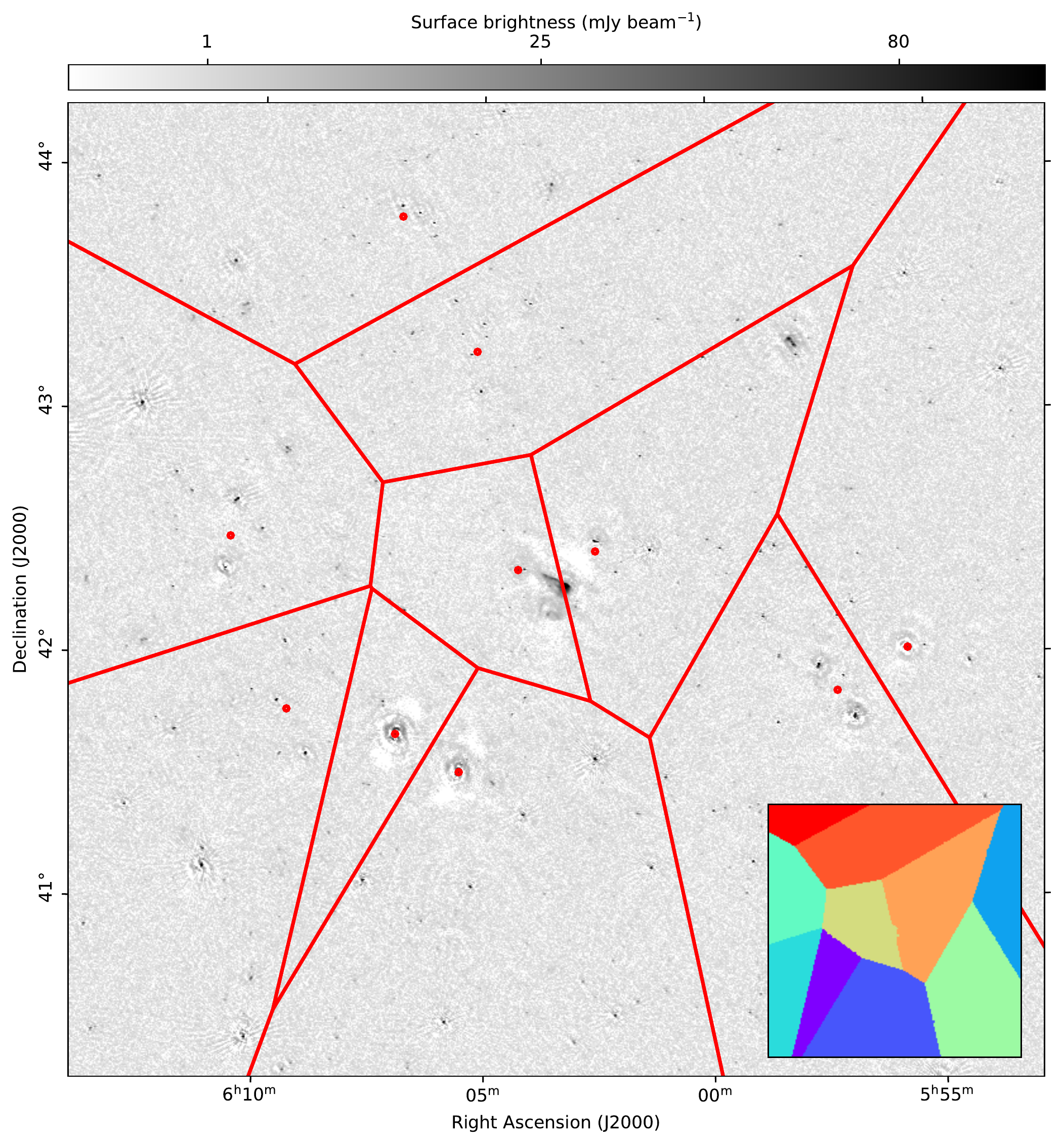}
 \caption{Radio map produced at the end of the direction-dependent calibration process. The rms noise is between 1.3 and 1.5~\mjybeam, depending on the image region, at a resolution of \beam{18}{11}. Red lines show the separation between facets and the red dots indicate the flux-weighted mean location of the sources used as direction-dependent calibrators. If a border between two facets crosses a source (as it happened for the primary target), the facet edge is modified so that each source is fully enclosed within one single facet. This is shown in the smaller panel where colour-coded there are the effective masks used to identify facets. We note that the border that goes over the extended emission of the Toothbrush cluster is not straight, so that the cluster is fully included in the left region.}
 \label{fig:map_dde}
\end{figure*}

The datasets are then corrected for the spatially averaged ionospheric corruption and sources outside an primary beam null are subtracted. The primary error remaining in the data are the severe differential direction-dependent errors caused the ionosphere. Without correcting these errors, high-fidelity imaging is not possible. To begin this procedure, sources in the model are grouped by proximity using a flux-weighted meanshift algorithm to isolate small clusters of sources. Extended sources and clusters with aggregated apparent flux density lower than 2 Jy are discarded as their flux would be insufficient to derive accurate solutions in those directions. The sources of the discarded clusters, together with all sources that were not included in any clusters, are subtracted from the dataset. We retained 10 clusters that we used to estimate the ionosphere corruption in those directions (see Fig.~\ref{fig:map_dde}). These clusters are called direction-dependent calibrators (DD-calibrators). Their aggregated flux density ranges between 2.4 and 9.5 Jy.

\begin{figure*}[!ht]
\centering
 \includegraphics[width=\textwidth]{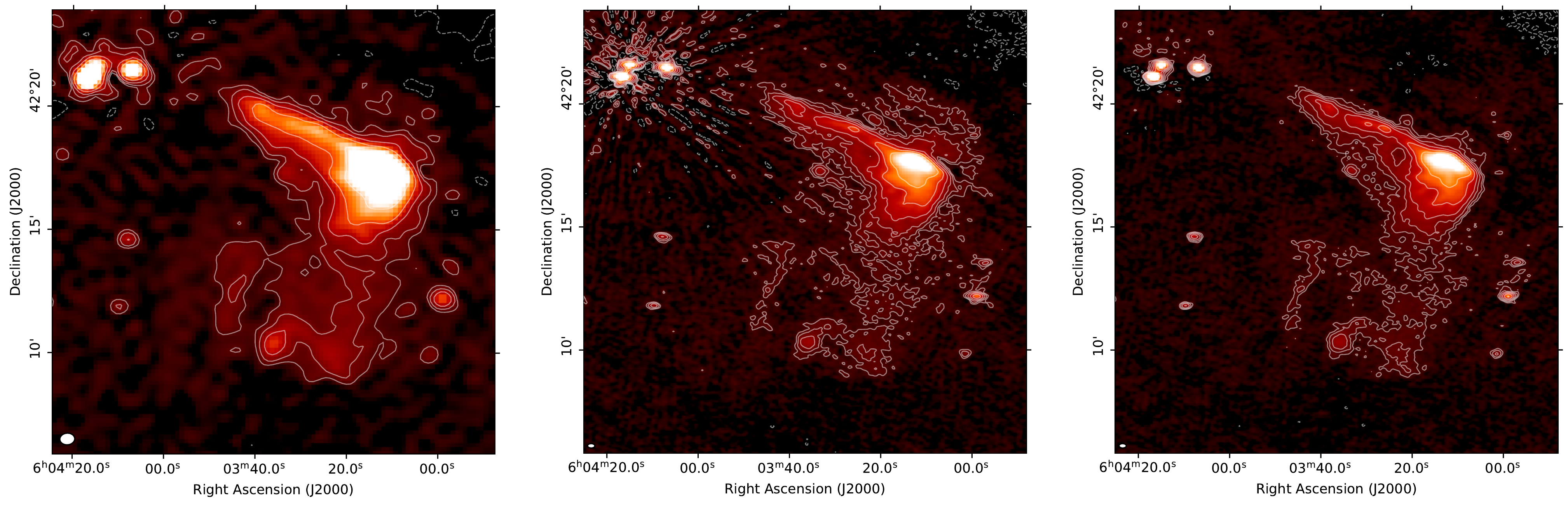}
 \caption{Left: Zoom-in on the Toothbrush cluster from Fig.~\ref{fig:map_die} (rms noise: $\sim 3$~\mjybeam, resolution: \beam{37}{29}). Center: Same dataset of the image of the left but re-imaged at resolution: \beam{18}{11}. Here, the effect of the ionospheric corruption mostly affecting long baselines is evident. Right: Final image after direction-dependent calibration and facet self-calibration (rms noise: 1.3~\mjybeam, resolution: \beam{18}{11}). In all images, the synthesised beam is shown in the bottom-left corner.}
 \label{fig:map_final}
\end{figure*}

For the direction-dependent solve, we used a simultaneous (for all relevant directions) scalar solver that enforces smoothness in frequency on scales of 5 MHz (Offringa et al. in prep.). The solve step is again done on smoothed data based on the baseline length, as in the case of the DIE self-calibration (see Sec.~\ref{sec:selfcal}). At this point we go back to the original dataset and subtract all sources while corrupting their expected visibilities with the ionosphere measured from the closest DD-calibrator. This produces a fully empty dataset, at least to the best of our ability given our ionospheric estimations. An obvious improvement over this strategy can come from the enforcement of spatial-coherency in the ionospheric TEC estimation \citep{Albert2020}. This will further increase the S/N and would improve the ionospheric correction in regions between DD-calibrators.

Finally, we used a Voronoi tessellation to create ten regions in the field of view (FoV), each related to one of the DD-calibrators. For each facet separately, all sources in the facet are added back to the empty dataset and the visibilities are corrected with the phases correction estimated for the associated DD-calibrator. Each facet is separately imaged at full resolution ($\sim 15\arcsec$). When all facets' maps are completed, they are combined to form a wide-field mosaic (Fig.~\ref{fig:map_dde}).

\subsection{Target extraction and self-calibration}
\label{sec:facet}

As a last step, we subtracted all sources from the dataset after corrupting their model for the phases estimated from the closest direction. We retained only around $0.5\deg \times 0.5\deg$ of sources around the target of interest (the Toothbrush cluster). A few cycles of scalar phase self-calibration was then performed with phase solutions obtained at increasingly higher time-resolution, from 32 down to 8 seconds.

In Fig.~\ref{fig:map_final}, we show the increase of the quality of the images going from the low-resolution image obtained after direction independent calibration to the high-resolution image (again, with only direction-independent correction) and up to the final high-resolution image after direction-dependent correction. The sequence of images shows very well how most of the ionospheric error is concentrated in the most remote stations and they are, therefore, not very visible in the first image. A direction-dependent calibration of the visibility phases is necessary to obtain a thermal noise, high-resolution image.

The final image (Fig.~\ref{fig:map_final}, bottom panel) has a noise of 1.3~\mjybeam{} and a resolution of \beam{18}{11}. To our knowledge, this is the deepest image ever obtained at these frequencies to date. The image was obtained using a Briggs weight of $-0.1$. The slightly higher value with respect to the expected thermal noise (1.1~\mjybeam) is likely due to the data loss due to interference and residual errors due to unmodelled flux in the side lobes and systematic effects. No primary beam correction has been applied as its variation in the region covered by the target is negligible.

\begin{figure}
\centering
 \includegraphics[width=.5\textwidth]{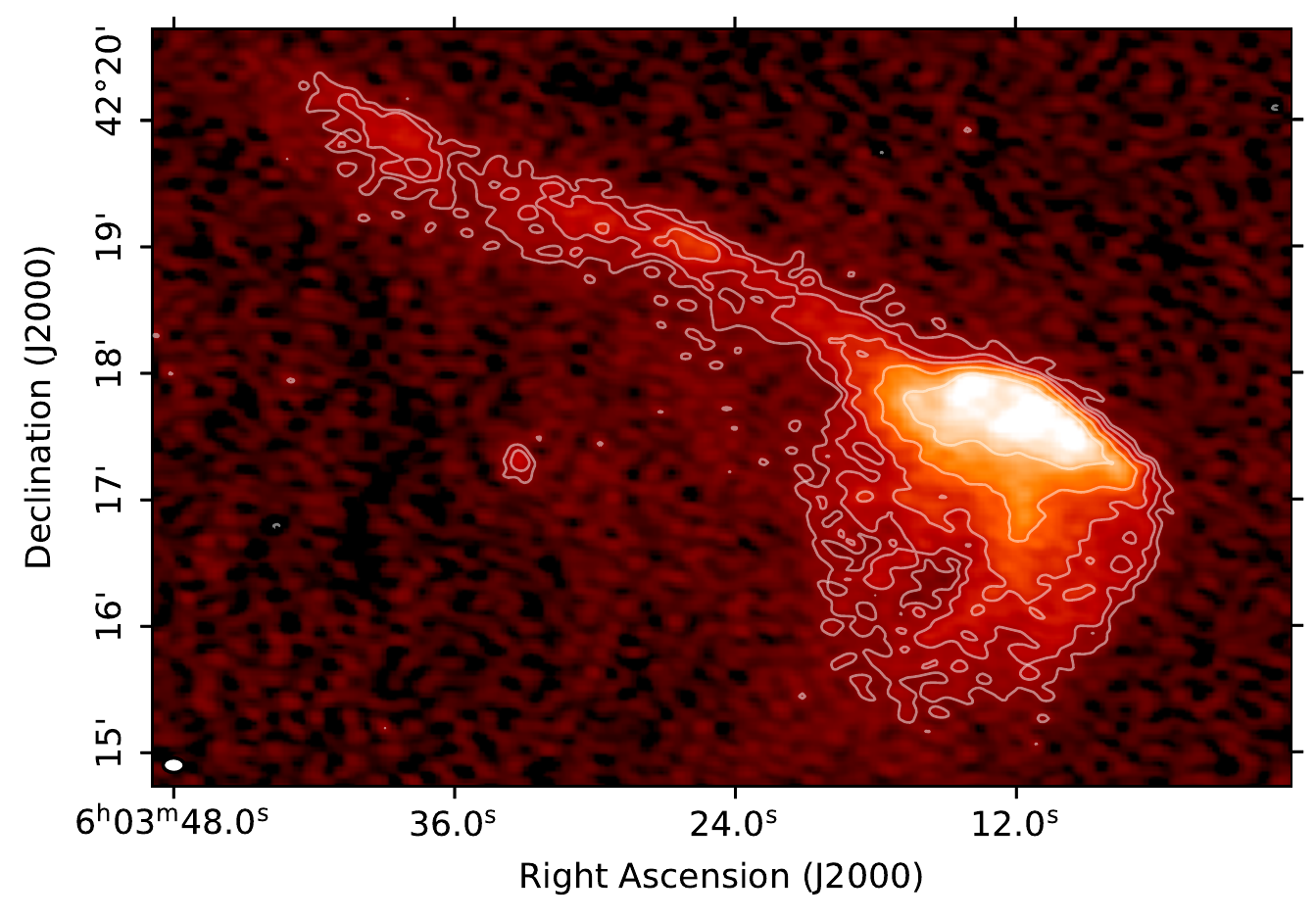}
 \caption{High-resolution image of the main radio relic at 58 MHz. With rms noise: 2.2~\mjybeam, resolution: \beam{10}{7} (beam shape in the bottom-left corner). Contour levels at: 8.8 ($4\sigma$), 14, 23, 38, 61, 100~\mjybeam.}
 \label{fig:map_final_hr}
\end{figure}

Finally, we also imaged the data at a higher resolution (see Fig.~\ref{fig:map_final_hr}) using a robust parameter of $-1$. This shows that  our direction-dependent calibration solutions are of high quality as we can still produce a high-fidelity, high-sensitivity (2.2~\mjybeam) image at a resolution of \beam{10}{7}.

\section{Results}
\label{sec:results}

\subsection{Spectral index maps}

\begin{figure*}
\centering
 \includegraphics[width=\textwidth]{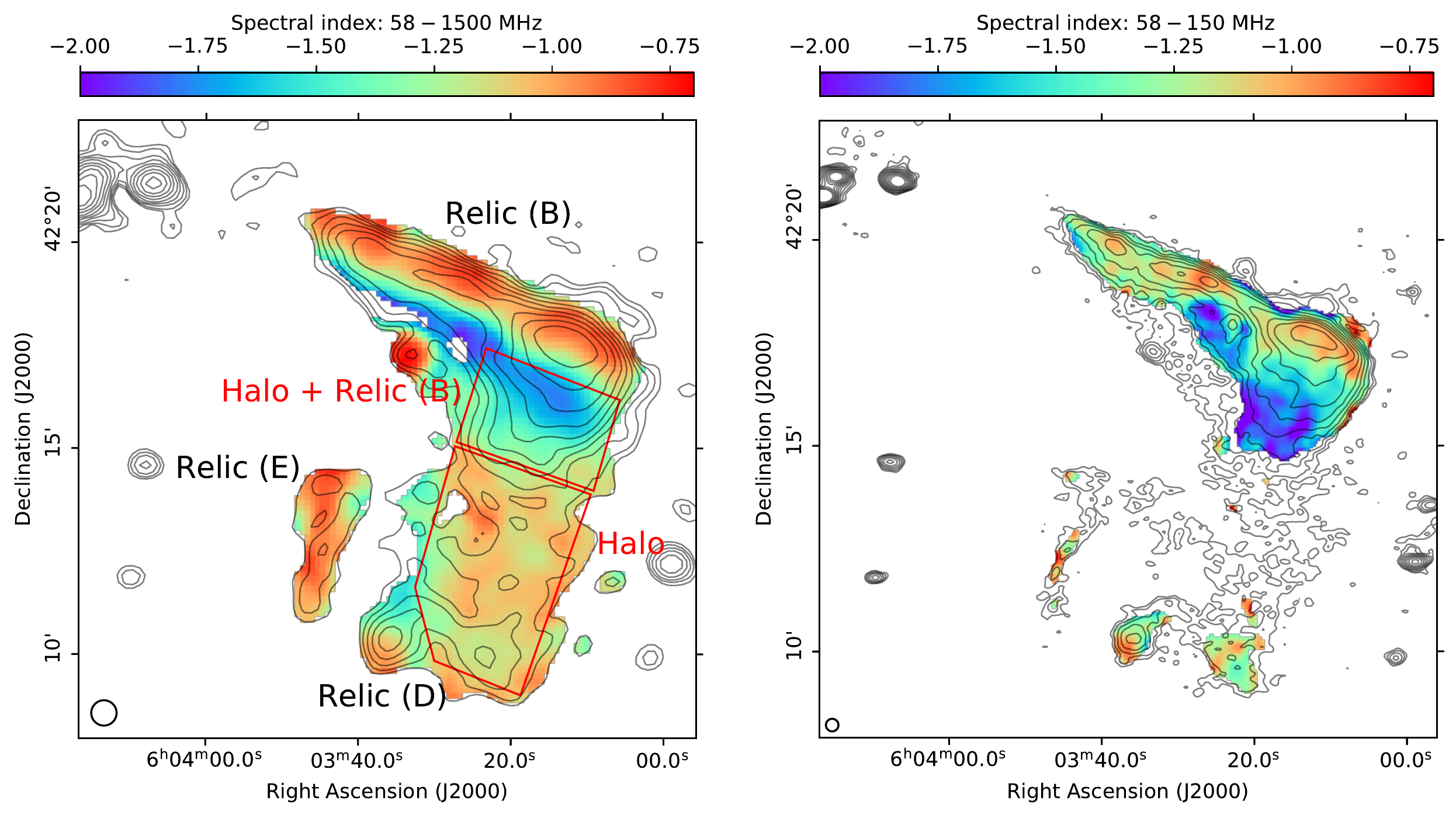}
 \caption{Left: low-resolution spectral index map 58 -- 1500 MHz (beam: \beam{37}{37}). Right: high-resolution spectral index map 58 -- 150 MHz (beam: \beam{18}{18}). Error maps are in Fig.~\ref{fig:spidx-err}. Labels as in \citep{vanWeeren2012e}.}
 \label{fig:spidx}
\end{figure*}

We used the dataset presented in this paper (58 MHz) to produce two spectral index maps: one making use of VLA observations (1500 MHz; Fig.~\ref{fig:spidx}, left panel) and another using LOFAR HBA observations (150 MHz; Fig.~\ref{fig:spidx}, right panel). Both VLA and LOFAR HBA datasets were presented in \cite{vanWeeren2016}. For details on the spectral index map calculation and error maps, see Appendix~\ref{sec:spidx}.

The first spectral index map is at lower resolution (37\arcsec) and covers both the radio relics and the halo region. To properly analyse this low surface brightness region, we used the widest possible frequency range in order to reduce the effect of spurious fluctuations. We extracted the spectrum from two regions (see Fig.~\ref{fig:spidx}): one covering the central-southern part of the halo (`Halo' region) and one covering the northern part of the halo (`Halo+Relic B' region).

In good agreement with results from \cite{vanWeeren2016} and \cite{Rajpurohit2018}, the integrated spectral index value of the halo region is $\alpha_{1500}^{58} = -1.10 \pm 0.05$ (`Halo region' in Fig.~\ref{fig:spidx}). The spectral index map in the Halo region does not show spectral trends, although we measure variations of the spectral index alpha of about $0.2-0.3$ that are significantly larger than the typical errors ($\leq 0.1$, see Fig.~\ref{fig:spidx-err}). A large portion of the halo emission overlaps with the emission coming from the aged cosmic ray (CR) electrons left behind by the merger shock wave (`Halo+Relic B region' in Fig.~\ref{fig:spidx}). This overlapping region has a steeper spectrum compared to the rest of the halo: $\alpha_{1500}^{58} = -1.54 \pm 0.05$. The spectral index map in the overlapping region shows a clear gradient thus the steeper spectrum is likely caused by the mix of the very steep spectrum emission downstream of the relic and the halo that are projected in the same area of the sky.

In the second panel of Fig.~\ref{fig:spidx}, we present a high-resolution (18\arcsec) low-frequency spectral index map obtained between 58 and 150 MHz. In this image, most of the extended emission of the radio halo has a surface brightness that is too low to be significantly detected and the radio relic emission becomes dominant. With this map, we can follow the ageing of the radio plasma for $\sim 1$~Mpc behind the shock front. Furthermore, the variation of the spectral index along the radio relic ridge becomes evident. 

\subsection{Large-scale spectral index analysis}

\begin{figure*}
\centering
 \includegraphics[width=.3\textwidth]{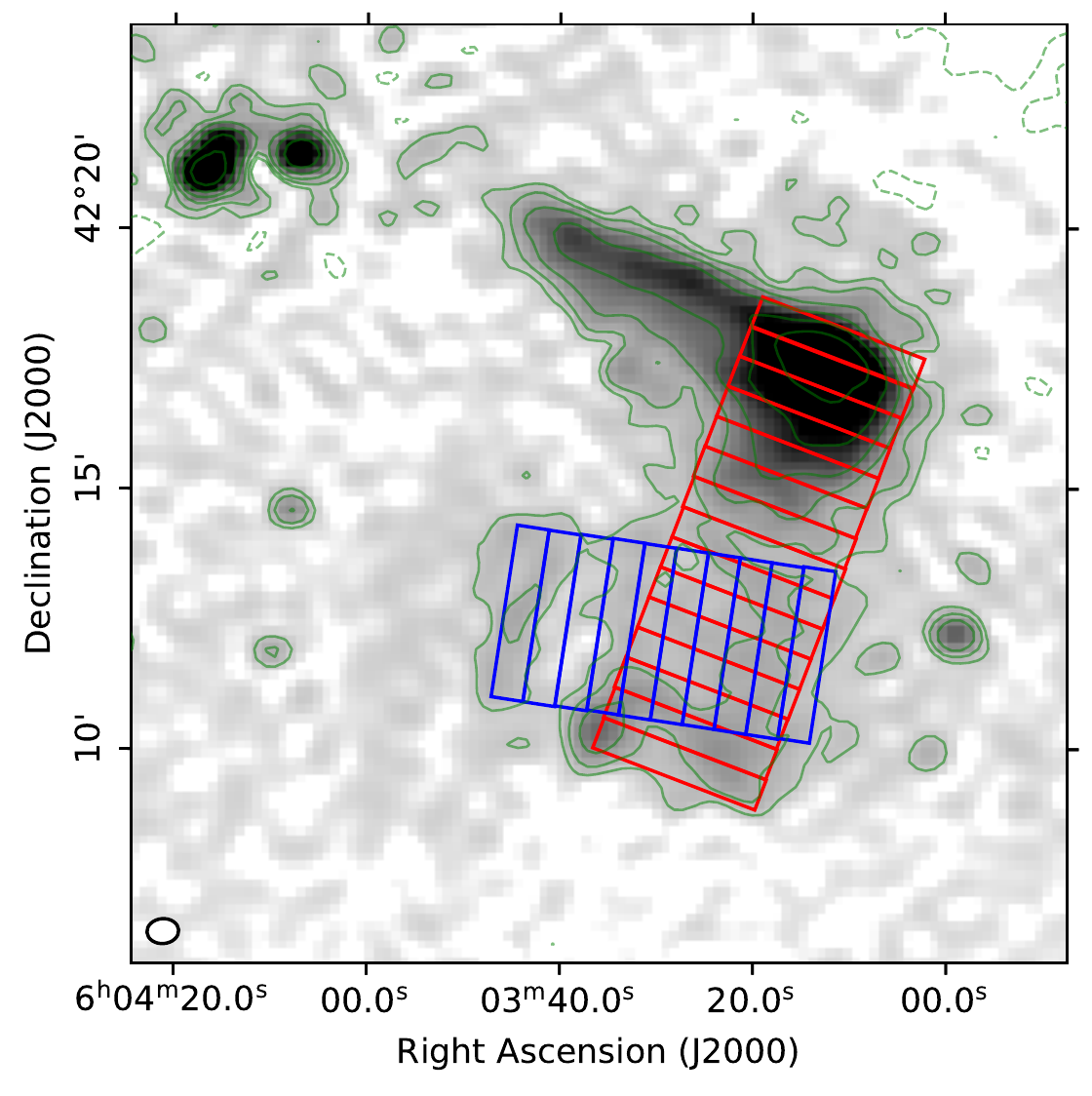}
 \includegraphics[width=.34\textwidth]{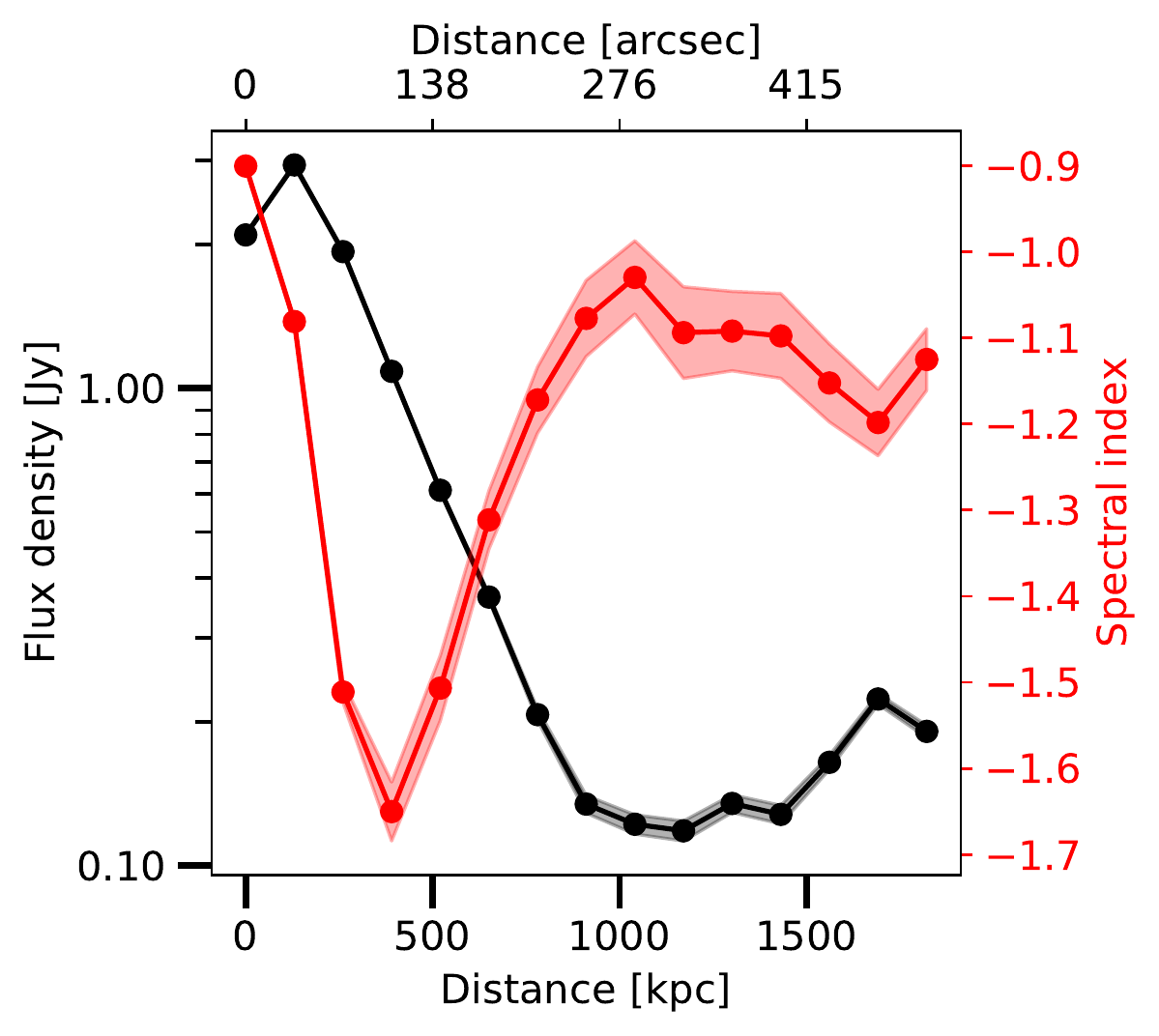}
 \includegraphics[width=.34\textwidth]{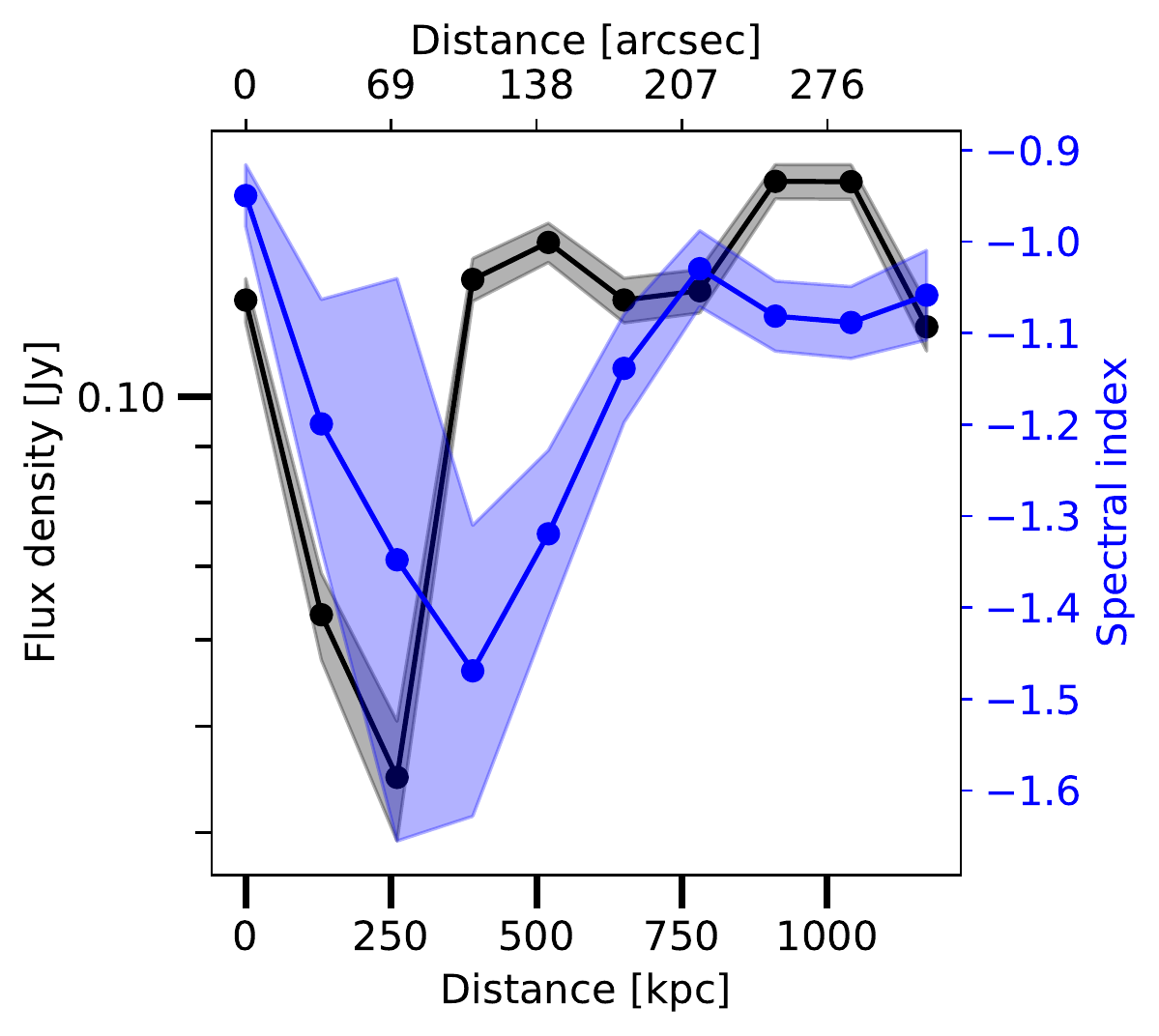}
 \caption{Left: Radio map at 58 MHz (same of Fig.~\ref{fig:map_final}, right panel) superimposed with the regions used to extract the flux density and the spectral index plotted in the central and right panels. The short side of the regions has the size of one beam (37\arcsec). Centre: Black represents the flux density at 58 MHz, and in red, the spectral index between $58-150$ MHz (dots) of the red regions (north $\rightarrow$ south). Right: Black represents the flux density at 58 MHz and in blue, the spectral index between $58-150$ MHz of the blue regions (east $\rightarrow$ west).}
 \label{fig:spidx_analysis_low}
\end{figure*}

To study the spectral index properties of the large scale emission, we used the two low-resolution radio maps at 58 and 1500 MHz which were previously used to extract the low-resolution spectral index map of Fig.~\ref{fig:spidx}. We selected two regions of the cluster where the two main radio relics are located. We divide the regions in beam-sized sub-regions so that we can measure the average flux density moving away from the relics, perpendicular to the shock front in the direction downstream of the shock, well into the radio halo. We show the regions in Fig.~\ref{fig:spidx_analysis_low} (left panel). In the case of the main radio relic (red regions), the flux density decreases departing from the shock front due to synchrotron and inverse Compton losses. This trend continues up to a distance of $\sim 800$~kpc where the surface brightness levels off to the level of the radio halo. Then it stays rather constant for around 1 Mpc. The spectral index instead decreases rapidly due to the quick depletion of energetic electrons. It reaches a minimum of $\alpha=-1.7$ at about 400 kpc from the shock front and then it rises again to reach the average halo spectral index ($\alpha \approx -1.1$) around 800 kpc away from the shock front.

In the case of the eastern relic (blue regions), the situation is similar. However, this relic is less powerful and the surface brightness at 58 MHz is comparable to that of the radio halo. Therefore, the emission decreases initially before increasing again in the halo region. The evolution of the spectral index is similar to that observed for the northern relic. However, contrary to the northern relic,  we note that the spectrum begins to steepen toward east direction when the emission of the halo is still dominant (Fig.~\ref{fig:spidx_analysis_low}, right), suggesting the presence of steep spectrum emission from the external region of the halo.

\subsection{High-resolution, low-frequency spectral index analysis}

\begin{figure*}
\centering
 \includegraphics[width=.3\textwidth]{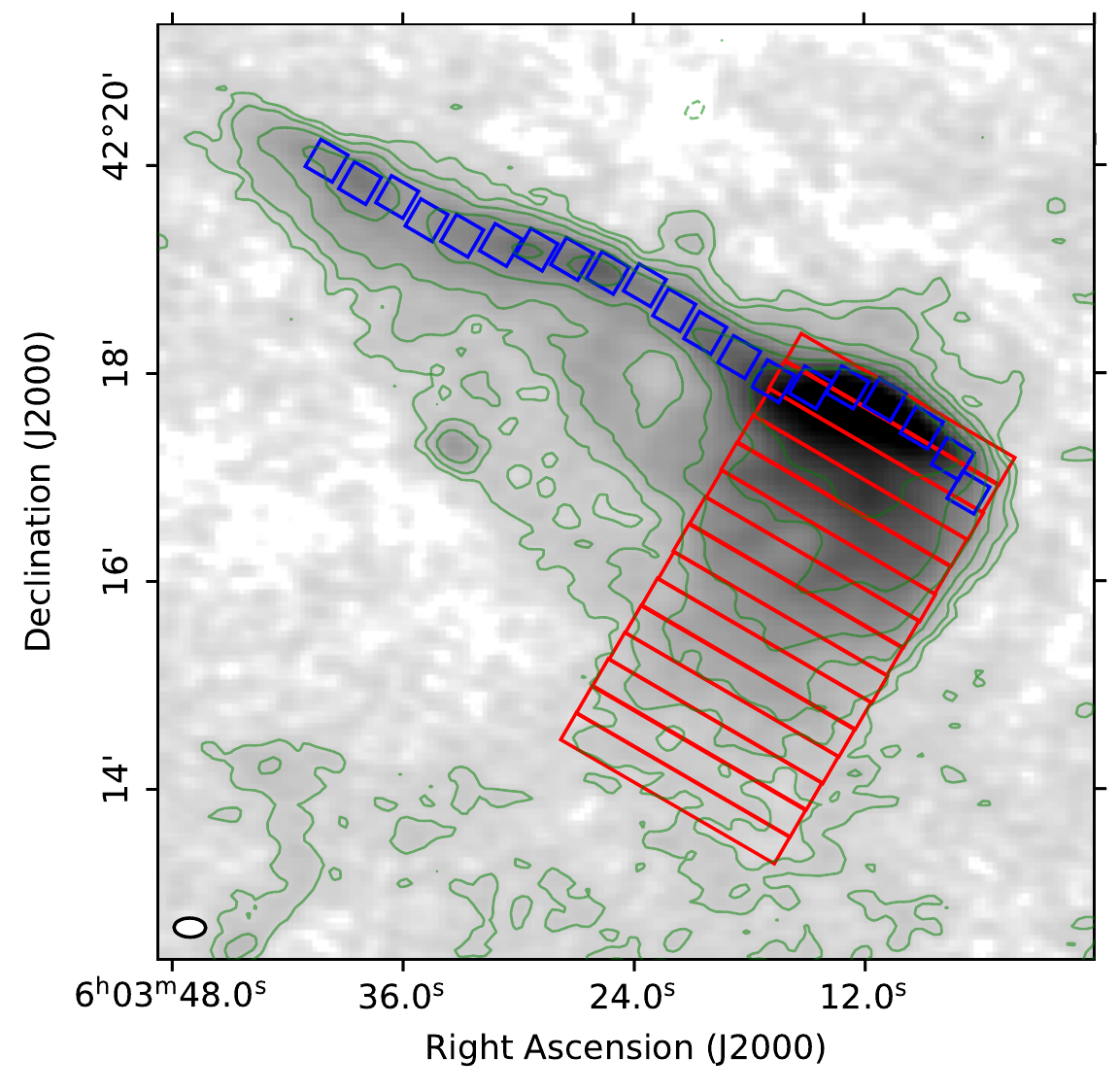}
 \includegraphics[width=.34\textwidth]{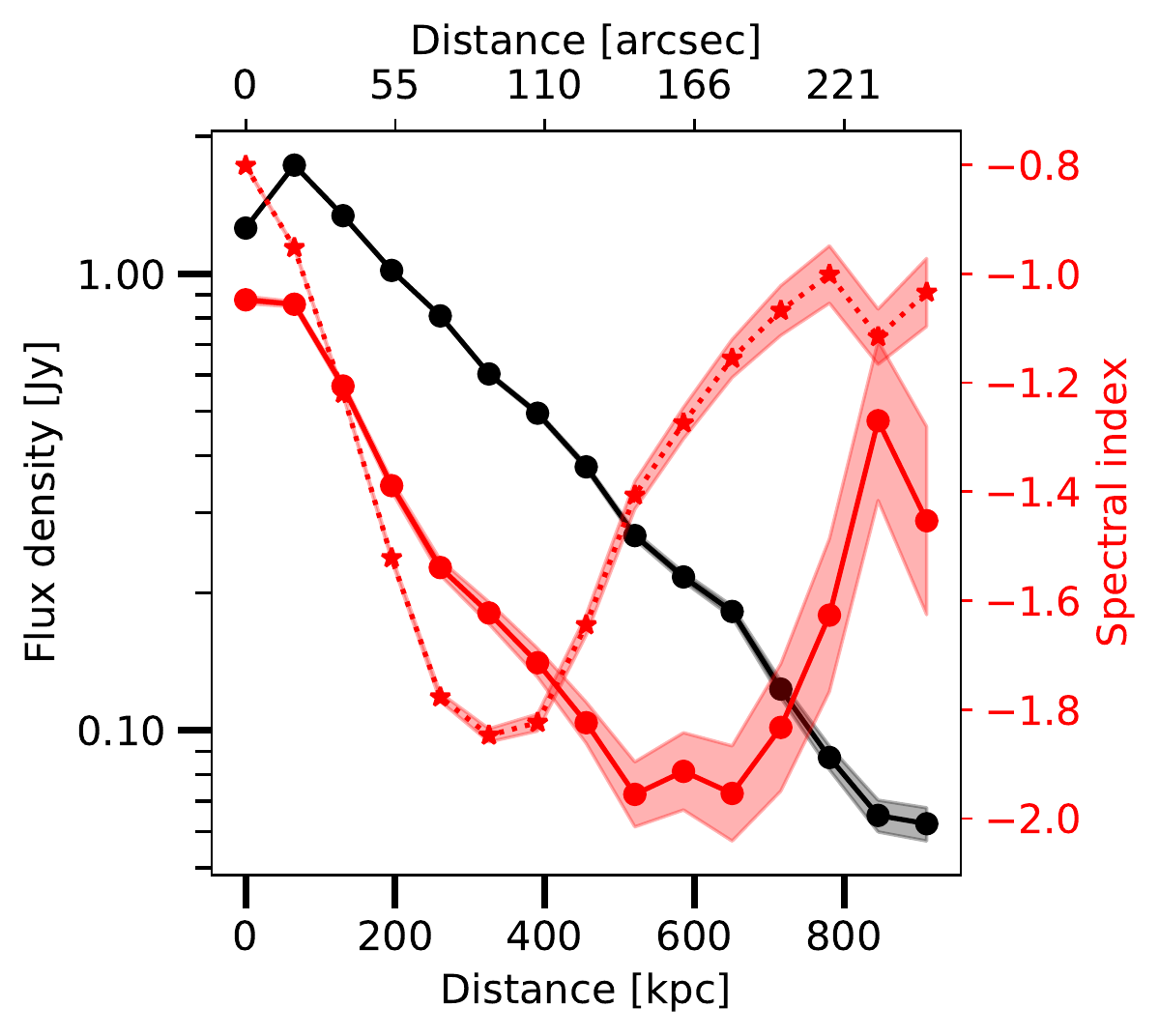}
 \includegraphics[width=.34\textwidth]{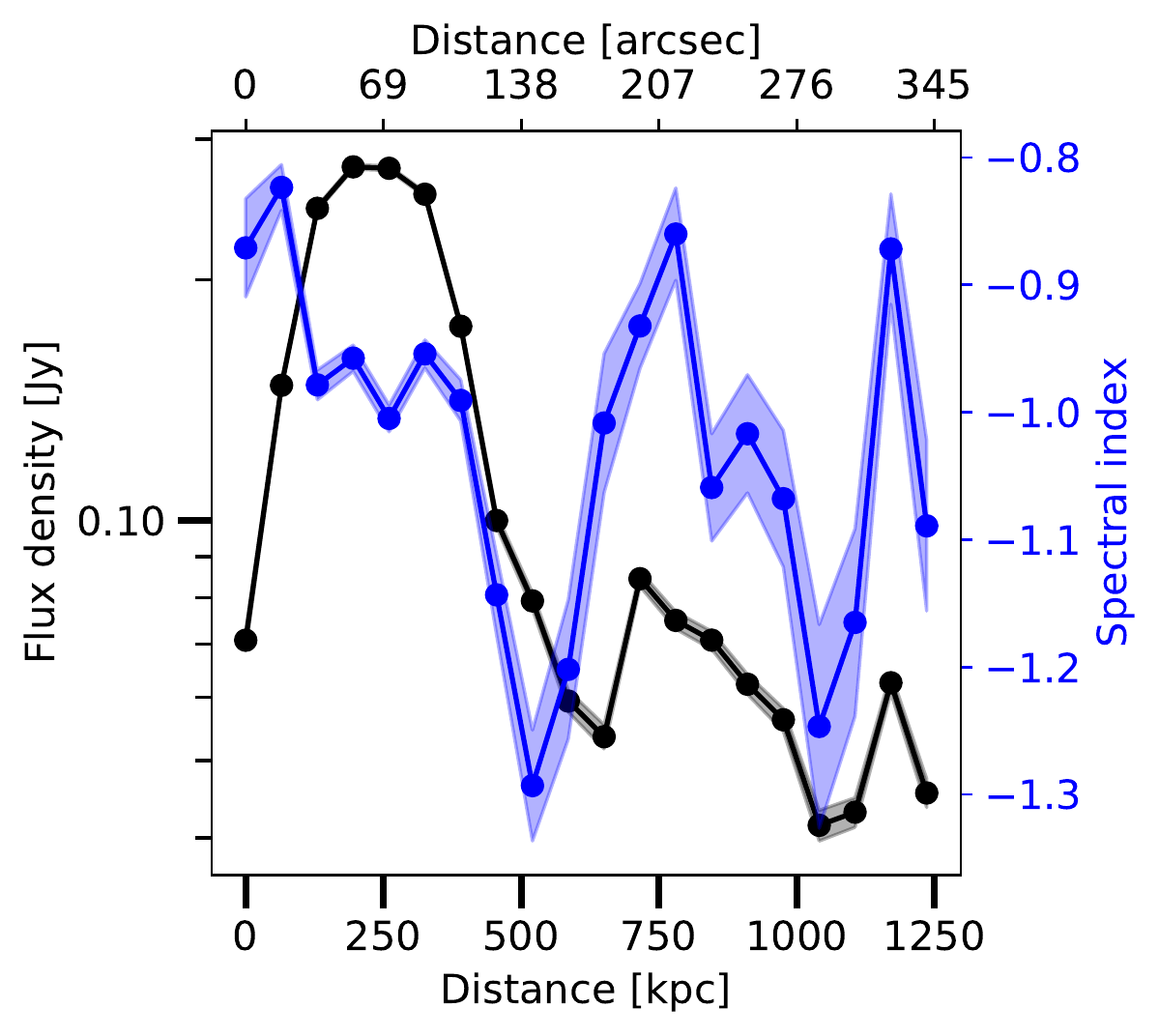}
 \caption{Left: Radio map at 58 MHz (same of Fig.~\ref{fig:map_final}, right panel) superimposed with the regions used to extract the flux density and the spectral index plotted in the central and right panels. Along the repeating direction, regions have the size of one beam (18\arcsec). Centre: Black represnts the flux density at 58 MHz and in red, the spectral index between $58-150$ MHz (dots) and $150-1500$ MHz (stars) of the red regions (north $\rightarrow$ south). Right: Black represents the flux density at 54 MHz, in blue, the spectral index between $58-150$ MHz of the blue regions (west $\rightarrow$ east).}
 \label{fig:spidx_analysis_high}
\end{figure*}

We also made $\alpha$-maps using higher resolution (18\arcsec) images at 58 and 150 MHz and estimated the spectral indexes along the brightest part of the radio relic B, going downstream from the shock front (see Fig.~\ref{fig:spidx_analysis_high}). We use these higher-resolution images to sample every 65 kpc and find that the average flux density decreases steadily until a distance of 800 kpc behind the shock front. At the same time, we can follow the decrease of the low-frequency spectral index from $\alpha \approx -0.8$ at the shock front down to $\alpha \approx -1.7$ at 600 kpc behind the shock front. This steepening is likely due to synchrotron and inverse Compton losses. Adiabatic losses should not be dominant as the ambient pressure is increasing with decreasing distance to the cluster centre. At the distance of 700 kpc, the spectral index starts to flatten again to finally reach $\alpha \approx -1.1$ around 800 kpc behind the shock front. This flattening of the spectrum is due to the presence of the radio halo that begins to dominate the emission at large distances from the shock. It is interesting to compare the changes of the high frequency spectral index ($150 - 1500$ MHz) in the same region (stars in Fig.~\ref{fig:spidx_analysis_high} - second panel). Here the turnover from a steepening to a flattening of the spectrum happens at $\approx 300$~kpc, well before the low frequency case.

Finally, we extracted the low-frequency spectral index along the relic edge, where the particles are accelerated by the shock wave. The spectral index values range between $\alpha = -0.5 \rightarrow -1.0$. In line with what has been seen by ~\cite{vanWeeren2016} and \cite{Rajpurohit2018}, the spectral index value is correlated with the flux density of the emitting region, with flatter spectra corresponding to brighter regions. An increase in the local magnetic field strength would increase the emissivity of the CR electrons, therefore sampling particles at lower energies for a given emitting frequency. If the electron energy distribution is curved, the net outcome would be a flattening of the spectral index. However, close to the shock front, the spectrum is expected to be straight and no change in spectral index would be expected. Alternatively, the observed variation of the spectral index along the relic extension can be linked to a complex shock surface. Simulations suggest that the shock surface can be composed by multiple shock fronts with a varying Mach number \citep{Skillman2013, Ha2018}. Observations also confirm that radio relics can be powered by shocks with non-uniform Mach number  \citep{deGasperin2015b}. As suggested by \cite{Hong2015}, a mix of shock surfaces with different Mach numbers associated with projection effects can also reduce the large discrepancy found in the Toothbrush cluster between the synchrotron weighted Mach number and the X-ray weighted Mach number, with the latter being smaller \citep{vanWeeren2016}. However, according to standard DSA of thermal electrons, the Mach number thus found still cannot explain the observed synchrotron luminosity due to the small amount of energy dissipated into particle acceleration at cluster shocks \citep{Botteon2020}.

\section{Discussion}
\label{sec:discussion}

\subsection{Possibility of re-accelerated CRs in the post-shock region}

Synchrotron emission at low frequencies is produced by low-energy electrons whose lifetimes are very long. For example, for the magnetic field that minimise losses of the electrons emitting at a given frequency, $B=B_{CMB}/\sqrt{3}=1.4~\mu G$, the lifetime of CR electrons emitting at 58 MHz is $\sim 290$ Myr. Thus, at lower frequencies, we can trace electrons farther downstream from the shock. In the case of the Toothbrush, we can detect CR electrons that we associate with the shock wave at distances of about 800 kpc from the shock. Assuming a downstream bulk velocity of 1000~\kms{}, this suggests that the emitting electrons at 800 kpc from the shock have been accelerated about $750-800$ Myr before those at the shock front and this is too long compared to their radiative lifetime. Specifically, \citet[][eq. 1]{Kang2017a} provides an estimation of the length of the post-shock region for the Toothbrush cluster at a certain observing frequency assuming simple synchrotron and inverse Compton ageing. The maximum expected width of the relic is expected to be $\Delta l_{58\rm\ MHz} = 211$~kpc, while we measure $\sim 800$~kpc. \cite{Kang2017a} already noted a discrepancy of a factor two with data at 610 MHz, but it becomes rather large (factor four) with our observations at 58 MHz. To overcome the problem, they proposed re-acceleration by post-shock turbulence.

One possibility could be that electrons are re-accelerated when entering in a vast turbulent region downstream of the shock where the radio halo is generated \citep[e.g.][]{Markevitch2005, Markevitch2012}. However, in looking at the second panel of Fig.~\ref{fig:spidx_analysis_high}, we can see that the putative re-acceleration seems to kick in around 300 kpc behind the shock. At the same time, the surface brightness of the 58 MHz emission, tracking the less energetic electrons, seems to continue to gradually decrease until 800 kpc downstream of the shock front. If electrons 'enter' into the halo region to be re-accelerated, the position where the steepening stops should be almost independent of frequency. This situation clearly suggests that the steeper component downstream of the relic is 'projected' onto the halo and not  'mixed' with the halo. So, going to lower frequencies, the steeper component gets brighter and it dominates the emission to larger distances.

This view has also been proposed in \cite{Rajpurohit2020}, where the radio relic is positioned in projection with respect to the radio halo. In this case, the two plasma populations do not mix at all and the visual effect of the gradual steepening and flattening in Fig.~\ref{fig:spidx_analysis_high} is just due to the line-of-sight superimposition of the two particle populations: the ageing tail of the radio relic and the flatter, faint radio halo. In any case, the process powering the radio halo seems unrelated to the one prolonging the life of CRs downstream the shock front, although both might be due to Fermi-II type acceleration.

\subsection{Estimating when the shock starts to accelerate CRs}

Looking at the flux density decrease along the relic ageing tail, we can confirm that the shock was already pumping energy into CR acceleration to roughly 800~kpc from the shock front. This is a location surprisingly close to the center of the cluster, which is $\approx 1$~Mpc from the shock front.

\begin{figure}
\centering
 \includegraphics[width=\columnwidth]{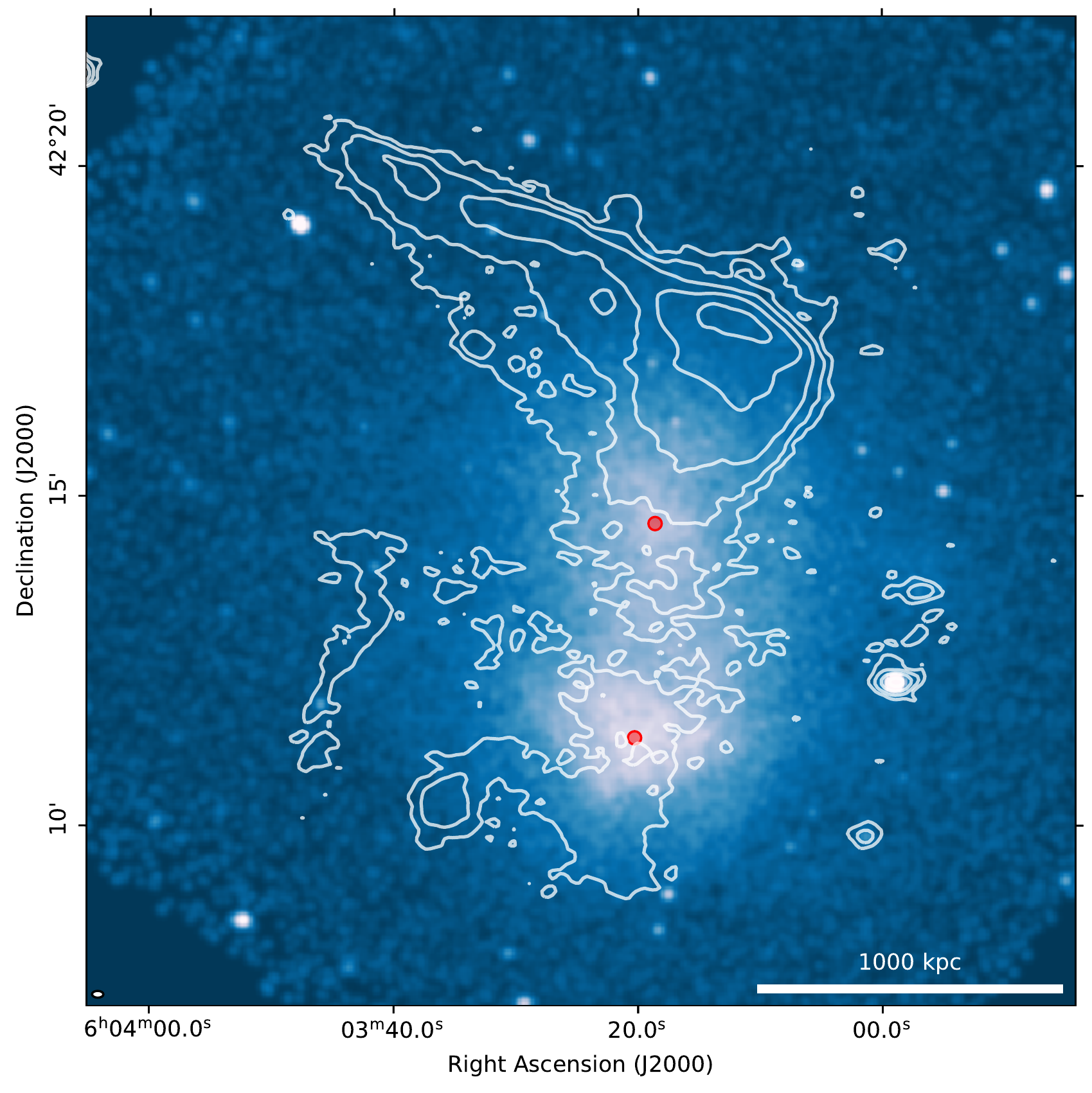}
 \caption{\textit{Chandra} 0.5--2.0 keV image with superimposed contours from Fig.~\ref{fig:map_final}. Red dots show the position of the two X-ray peaks. The emission directly related with the radio relics arrives $\sim 260$ kpc north of the mid-distance between the two markers.} \label{fig:xray}
\end{figure}

We used the Chandra image published in \cite{vanWeeren2016} to locate the mid-distance between the two X-ray peaks associated with the two merging sub-clusters (see Fig.~\ref{fig:xray}). We take this as a rough estimation of the crossing point of the two clusters at merging time. The end point of the radio relic downstream tail is 72\arcsec{} ($\approx 260$~kpc) away from the crossing point along the line connecting the two X-ray peaks. The presence of multiple radio relics and of multiple X-ray peaks suggests that the merger happened in a plane close to the plane of the sky. This is challenged by recent colour-colour measurements on the northern relic that suggests some level of projection \citep{Rajpurohit2020}. Assuming an inclination of 30\deg{} between the shock propagation direction and the plane of the sky, the distance where the first particles are detected goes up to 300~kpc. This is most likely an upper limit as, once shock accelerated, the CR particles should retain some momentum in the shock moving direction.

Radio relics are extremely rare at these short distances (see \cite{VanWeeren2009b, Bonafede2012a, deGasperin2014c}, also see \cite{Vazza2012} for a discussion). The fact that we see relic-related emission generated so close to the cluster center in the Toothbrush cluster can be related to the high brightness of this specific radio relic\footnote{The Toothbrush radio relic is the most powerful radio relic known, see e.g. \cite{vanWeeren2019}.} or to a rather important projection effect. Furthermore, compared to other radio relics such as the `Sausage' radio relic in CIZA J2242.8+5301 \citep[see e.g.][]{Hoang2017}, the Toothbrush radio relic has filamentary streams in the downstream region, oriented perpendicularly to the shock front. These streams might be the result of stretched magnetic field lines that allow electrons to diffuse more efficiently along these field lines \citep{vanWeeren2016}. It is currently unclear why the Sausage and the Toothbrush downstream regions are so different but observations at ultra-low frequencies ($10-100$~MHz) can help to shed some light on this aspect.

\subsection{The low-frequency spectral index of the radio halo}
\label{sec:models}

The integrated low-frequency spectral index of the radio halo is $\alpha_{150}^{58} = -1.16 \pm 0.15$ (`Halo' region in Fig.~\ref{fig:spidx}). This is in line with the spectral index found between VLA data at 1500 MHz and LOFAR HBA data at 150 MHz \citep[$\alpha_{1500}^{150} = -1.17 \pm 0.04$][]{Rajpurohit2018}. This shows that even when going down to the lowest frequencies, the spectral shape of the radio halo does not deviate from a power law. A steepening of the halo spectrum going at higher frequency has been predicted by the turbulent re-acceleration theory \citep{Cassano2006, Cassano2012}. This has been indeed noticed in the halo in the Coma cluster, whose spectrum steepens from $-1.2$ to $-2.3$ above 1500 MHz \citep{Thierbach2003}. However, the degree of curvature depends on the physical conditions. For homogeneous models with the same acceleration efficiency in the volume, simultaneous acceleration of all particles, and constant magnetic field in the volume, the curvature is stronger. Departure from homogeneity across the volume stretch the spectrum making any existing curvature less evident \citep[see e.g. simulations by][]{Donnert2013}. The fluctuations observed in the spectral index map qualitatively support a departure from homogeneous conditions in the halo volume. If confirmed by deeper observations, these fluctuations combined with the evidence of a power law spectrum extending across a large frequency range provides a challenge for spectral models. Furthermore, we also measure a steepening in the spectrum of the radio halo in the direction of the E-relic. This steepening was noticed also at higher frequencies \citep{vanWeeren2016} and may be caused by a decline of the magnetic field in the external region of the halo or by a steeper spectrum of the accelerated electrons \citep[e.g.][]{Brunetti2001}. 

It is important to note that at 58 MHz, around a third of the radio halo extension (as measured at 1500 MHz) is polluted by the steep spectrum emission from electrons downstream of the relic.
Contamination between the two components underline the intrinsic difficulties in isolating emission coming from different acceleration processes when analysing ultra-low frequency radio maps of galaxy clusters. For example, \cite{vanWeeren2012f} detected a steepening of the spectrum at low frequencies in Abell 2256. This steepening might be explained by the contamination between halo and relic (or radio galaxies). From Fig.~\ref{fig:spidx}, the presence of aged electrons is clearly visible in the north part of the halo, region 'Halo + Relic (B)', but also to the south of the halo, close to 'Relic (D)'.

\subsection{A shock model with fossil CRe}

Our observations at ultra low frequency have shown an extension of the relic that is in conflict with a scenario based on the ageing of electrons downstream. Here, we sketch two possible models to explain the profile of the spectral index shown in Fig.~\ref{fig:spidx_analysis_high}. 

In the first scenario, we assume that the electrons downstream in the relic can be maintained at higher energies by re-acceleration mechanisms for a time that is much longer than their cooling time. This model has been already proposed for the relic in the toothbrush by \citet{Kang2017} to explain the observations at 150 and 610 MHz. Specifically, they assumed that during the merger a shock emerged near the cluster merger about 800 Myr ago and that it encountered a extended cloud of fossil CRe and turbulent magnetic field. According to \citet{Kang2019} in supercritical, quasi-perpendicular shocks with $\mach \geq 2.3$, micro-instabilities, such as the electron firehose instability, may self-generate electron scale magnetic turbulence that can provide necessary upstream scattering waves. \citet{Kang2017} assumed downstream acceleration from shock-excited turbulence and showed that a model assuming $\mach \sim 3$ can reproduce the spatial profile of the spectral index observed in the toothbrush relic up to 200 kpc distance downstream (see their Fig.~4). The turbulent acceleration model adopted by \citet{Kang2017} assumes transit-time-damping resonance with compressive shock-excited post-shock turbulence that decays downstream with a length scale of $\sim 100$ kpc. However, our observations require that turbulence re-accelerates electrons for much longer times and distances downstream. We propose that the additional acceleration on longer times and distances (200-800 kpc behind the shock) may be provided by background magnetic turbulence in the cloud of fossil relativistic CRe. Our proposal is motivated by the fact that the cloud of fossil plasma could originate from an extinct FRI jet, which has entrained the ICM gas and undergone turbulent mixing \citep[e.g.,][]{Laing2014}, providing turbulence which may linger even after the shock-generated turbulence has decayed.

\begin{figure}
\centering
 \includegraphics[width=\columnwidth]{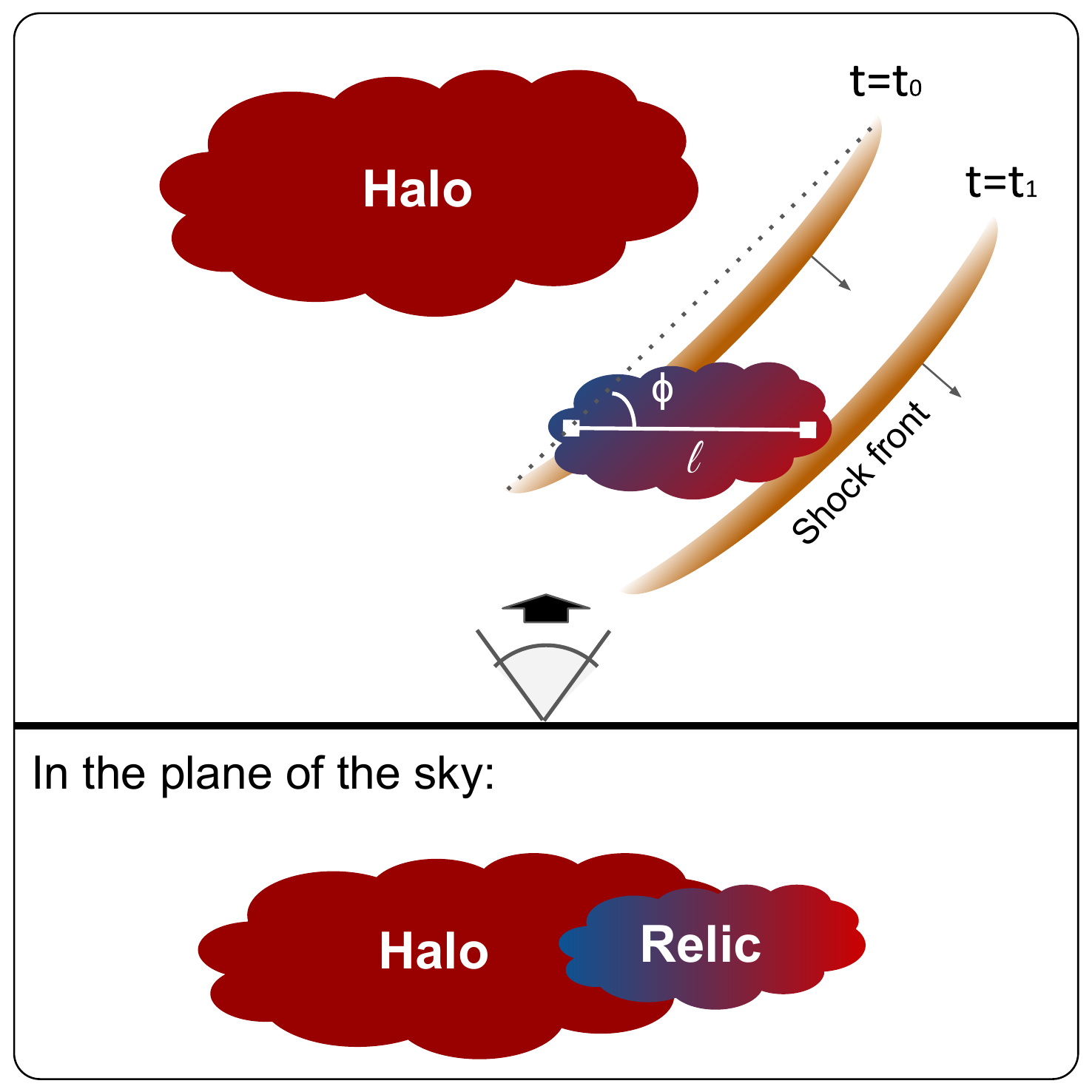}
 \caption{Figure represent a possible scenario explained in the second part of Sec.~\ref{sec:models}. Here the top panel shows the cluster as seen in projection, while the bottom panel shows the cluster as it appears in the plane of the sky. We show the shock front in two moments, when it first encounter the cloud of aged plasma ($t_0$) and in the current position ($t_1$). The colour encodes the radio spectral index of the plasma from red (flat) to blue (steep), as it is expected to be at $t=t_1$.} \label{fig:cartoon}
\end{figure}

An alternative picture that overcomes the requirement of a 800 kpc large turbulent cloud of pre-existing CRe is based on projection effects. If the shocks surface makes a small angle, $\phi,$ with a cloud of fossil CRe it is possible that the effective shock crossing time is much shorter than that estimated from the observed projected size of the relic downstream. This is shown in Fig.~\ref{fig:cartoon}, where we assume that the cloud of fossil electrons lies in the periphery of the cluster displaced along the line of sight. In this case, neglecting the thickness of the cloud, the effective shock crossing time is:


\begin{equation}
\tau \sim \frac{\ell \sin \phi}{c_{s,u}} = 
\frac{\ell}{c_{s,d}} \frac{\sqrt{(5\mach^2-1)(\mach^2+3)}}{4\mach^2} \sin{\phi}.
\end{equation}

with $c_{s,u|d}$ the upstream/downstream sound speed. Using the observed $\ell \sim 800$ kpc and assuming $c_{s,d}~1500$~\kms{} and $\mach \sim 2.5$, we find that the effective shock crossing time becomes shorter than the electrons cooling time, with $\tau \sim 200-300$ Myr for an angle $\phi < 30\deg$, thus removing the inconsistency between crossing time and cooling time of electrons downstream. We also note that in this scenario the shock is moving at a large angle with the plane of the sky providing a simple explanation to the fact that X-ray observations failed in detecting a clear shock at the position of the radio relic.

The two models discussed in this section are devoted to explain the extension downstream of the western part of the relic. The morphology of the entire relic with a fairly thin eastern region (i.e. the handle of the brush) and a very extended western region remains a mystery.

\section{Conclusions}
\label{sec:conclusions}

In this paper, we present a procedure to obtain thermal-noise limited images at ultra low-frequency ($<100$~MHz) with the LOFAR LBA system. We demonstrated our procedure using an observation of the Toothbrush galaxy cluster centered at 58 MHz, producing a final image with 1.3~\mjybeam{} rms noise at a resolution of \beam{18}{11}. We showed that by increasing the weights on the longest baselines, we could reach a resolution of \beam{10}{7}, but at a reduced sensitivity of 2.2~\mjybeam{}.

We combined the 58 MHz image of the Toothbrush cluster with radio maps at higher frequencies (LOFAR HBA at 150 MHz and VLA at 1500 MHz) to perform large scale and low frequency spectral studies. These are our findings:

\begin{itemize}
 \item Based on the spectral analysis of the radio relic, we provide additional strong evidence that the western part of the relic, exhibiting a very steep spectrum, is projected on the radio halo rather than mixed with it.
 \item Our observations at low frequencies show that the radio relic was active already 800 kpc behind the current shock front. This location is rather close to the cluster center, where presumably the cores of the two merging cluster encountered. At these distances from the core, relics are extremely rare, suggesting that the downstream region of the relic in the toothbrush may be projected onto the central regions of the clusters. 
 \item Our low-frequency observations show an extension of the relic downstream that is too large given the cooling time of electrons. Models based on shock re-acceleration of a cloud of fossil electrons offer a chance for us to remove this discrepancy. 
 \item In particular, we discuss the possibility that electrons are re-accelerated downstream by background turbulence in the cloud or that projection effects play a role when looking at a shock that crossed a peripheral cloud at a large angle.
 \item We provide the lowest frequency point of the integrated radio spectrum of the radio halo in RX J0603.3+4214. The spectrum is a power law with slope $\alpha \sim -1.1$ between 58 MHz and 1500 MHz. This finding combined with the evidence of fluctuations in the spectral index map might suggest inhomogeneous conditions in the halo volume if we assume that the halo originates from turbulent re-acceleration of electrons.
\end{itemize}

\subsection{Next steps}
\label{sec:nextsteps}

Observing at ultra-low frequencies is still very challenging and the ionospheric conditions during the observation can make the difference between a usable or an unusable dataset. However, owing to the vicinity of the minimum of the Solar activity cycle and the improved performance of the LOFAR LBA system, we can now routinely image the 30-70 MHz sky, de facto opening the lowest part of the radio window to deep, high-resolution observations. This will have a large impact in many areas of radio astronomy where constraining the low-frequency emission is important, including, for example, the detection of radio emission from exoplanets \citep[e.g.][]{Vedantham2020}, the discovery of the most distant radio galaxies and quasars \citep[e.g.][]{Saxena2018}, the origin and properties of old diffuse radio emitting plasma in galaxy clusters and AGN \citep[e.g.][]{Mandal2020}, star formation in nearby galaxies \citep[e.g.][]{Heesen2018}, properties of supernova remnants and the interstellar medium through absorption \citep[e.g.][]{Arias2018} and radio-recombination lines \citep[e.g.][]{Emig2018}, and possibly the characterisation of new classes of astronomical objects. In our next steps, we are using the LOFAR LBA system to image wider areas (500 deg$^2$ from the LOFAR LBA sky survey; de Gasperin et al. in prep.), deeper (sub-mJy; Williams et al. in prep.), and to exploit the resolution of the international LOFAR stations \citep[e.g.][]{Morabito2016}. This effort will continue with the upgrade of LOFAR (the so-called LOFAR 2.0) that will enable the use of all LBA dipoles simultaneously as well as combined LBA+HBA observations. These capabilities will increase the survey speed of the LBA system by a factor $>2,$ enabling the first deep, wide-area survey at ultra-low (10 -- 90 MHz) frequencies.


\begin{acknowledgements}

LOFAR is the LOw Frequency ARray designed and constructed by ASTRON. It has observing, data processing, and data storage facilities in several countries, which are owned by various parties (each with their own funding sources), and are collectively operated by the ILT foundation under a joint scientific policy. The ILT resources have benefitted from the following recent major funding sources: CNRS-INSU, Observatoire de Paris and Universite d’Orleans, France; BMBF, MIWF-NRW, MPG, Germany; Science Foundation Ireland
(SFI), Department of Business, Enterprise and Innovation
(DBEI), Ireland; NWO, The Netherlands; The Science and
Technology Facilities Council, UK; Ministry of Science and Higher Education, Poland; Istituto Nazionale di Astrofisica (INAF).
This work is partly funded by the Deutsche Forschungsgemeinschaft under Germany's Excellence Strategy EXC 2121 ``Quantum Universe'' 390833306.
WLW acknowledges support from the ERC Advanced Investigator programme NewClusters 321271. WLW also acknowledges support from the CAS-NWO programme for radio astronomy with project number 629.001.024, which is financed by the Netherlands Organisation for Scientific Research (NWO). HK was supported by the National Research Foundation of Korea (NRF) through grants 2016R1A5A1013277 and 2017R1D1A1A09000567. GB acknowledges partial support from INAF mainstream program "GALAXY CLUSTER SCIENCE WITH LOFAR". AB and RJvW acknowledge support from the VIDI research programme with project number 639.042.729, which is financed by the Netherlands Organisation for Scientific Research (NWO).
\end{acknowledgements}


\bibliographystyle{aa}
\bibliography{library}


\onecolumn
\begin{appendix}

\section{Spectral index maps}
\label{sec:spidx}

\begin{figure*}[h]
\centering
 \includegraphics[width=\textwidth]{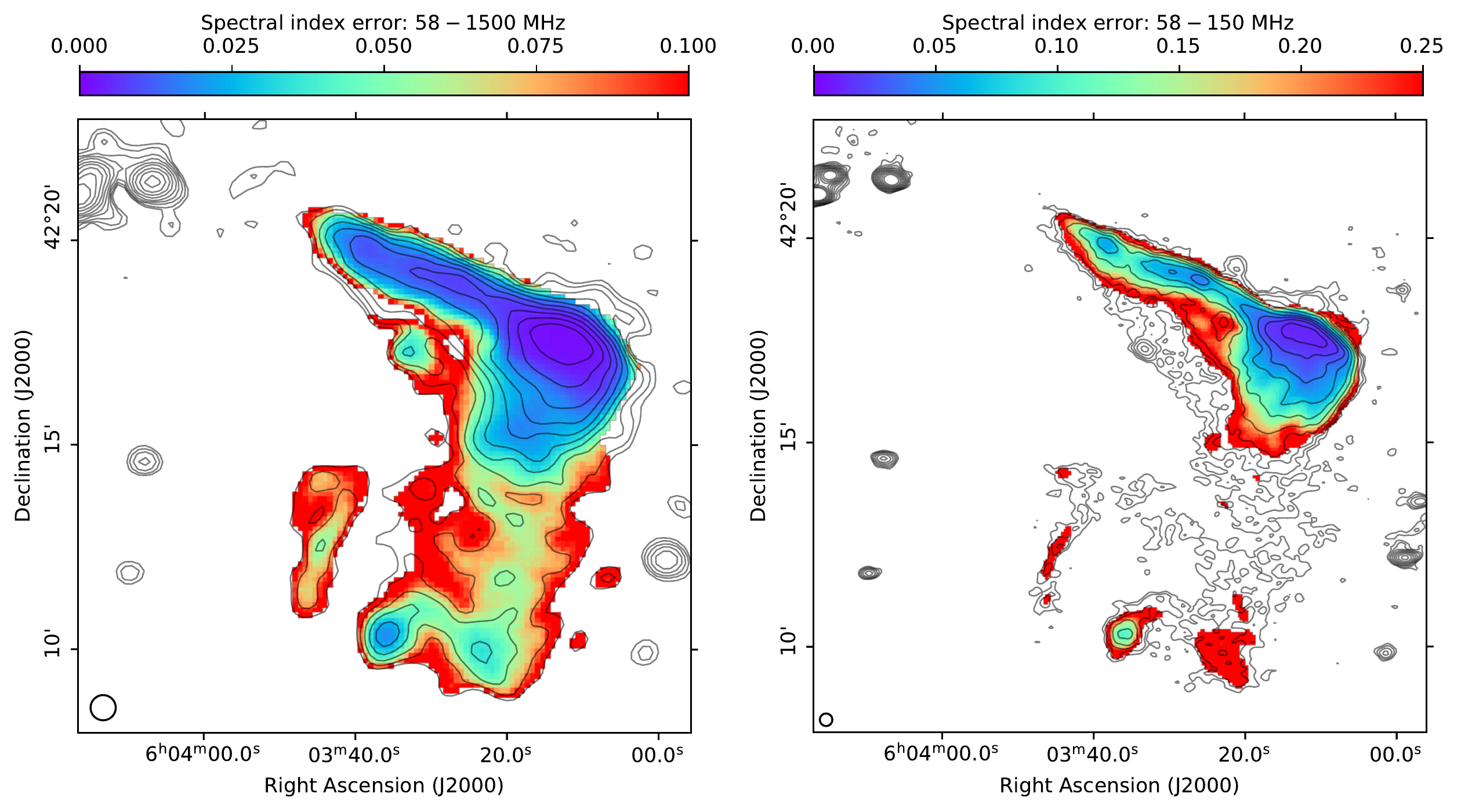}
 \caption{Spectral index error maps for the two panels in Fig.~\ref{fig:spidx}.}
 \label{fig:spidx-err}
\end{figure*}

The spectral index map in Fig.~\ref{fig:spidx} are produced by combining two images. For each image, each pixel with values below $3\sigma$ (with $\sigma$ being the rms noise of the image) is blanked. Flux density values are extracted 1000 times from a Gaussian distribution that has as the mean value the estimated flux density and as standard deviation the rms noise. The flux densities are combined to estimate the spectral index using:
\begin{equation}
 \alpha = \left. \log\left( \frac{S_{\nu1}}{S_{\nu2}} \right)\ \middle/\ \log\left( \frac{\nu_1}{\nu_2} \right) \right.,
\end{equation}
where $S_\nu$ is the flux density at the frequency $\nu$. The error on each spectral index value is estimated from the standard deviation of the 1000 values extracted. These errors are shown in Fig.~\ref{fig:spidx-err}.

\end{appendix}


\end{document}